\documentclass[11pt,a4paper]{article}

\usepackage{jheppub}
\usepackage{bm}
\usepackage{appendix}
\usepackage{latexsym}
\usepackage{verbatim}

\newcommand{\Dp}{D$p\,$}
\newcommand{\Dppf}{D$(p+4)$}
\newcommand{\PP}[2]{ \frac{\partial {#1}}{\partial {#2}} }
\newcommand{\DD}[2]{ \frac{d{#1}}{d{#2}} }
\newcommand{\IEH}{ I_{\rm EH} }
\newcommand{\IB}{ I_{\rm boundary} }
\newcommand{\IGH}{ I_{\rm GH} }
\newcommand{\intx}{ \int d^{10}x }

\newcommand{\ointx}{ \oint d^9x }

\newcommand{\eval}[1]{ \,\rule[-10pt]{0.4pt}{25pt}_{\,#1} }
\newcommand{\Q}{ {\cal Q} }
\newcommand{\TQ}{ \tilde{\cal Q} }

\newcommand{\DP}{ \Delta_+ }
\newcommand{\DM}{ \Delta_- }
\newcommand{\DS}{ \Delta_* }
\newcommand{\RP}{ \rho_+ }
\newcommand{\RM}{ \rho_- }

\newcommand{\BRP}{ \bar{\rho}_+ }
\newcommand{\BRM}{ \bar{\rho}_- }

\newcommand{\bphi}{ \bar{\phi} }
\newcommand{\brho}{ \bar{\rho} }
\newcommand{\bbeta}{ \bar{\beta} }
\newcommand{\BDP}{ \bar{\Delta}_+ }
\newcommand{\BDM}{ \bar{\Delta}_- }
\newcommand{\BDS}{ \bar{\Delta}_* }
\newcommand{\ICC}{ I_{\scriptscriptstyle CC} }
\newcommand{\ICG}{ I_{\scriptscriptstyle CG} }
\newcommand{\IGC}{ I_{\scriptscriptstyle GC} }
\newcommand{\IGG}{ I_{\scriptscriptstyle GG} }
\newcommand{\RICC}{ \tilde{I}_{\scriptscriptstyle CC} }
\newcommand{\RICG}{ \tilde{I}_{\scriptscriptstyle CG} }
\newcommand{\RIGC}{ \tilde{I}_{\scriptscriptstyle GC} }
\newcommand{\RIGG}{ \tilde{I}_{\scriptscriptstyle GG} }
\newcommand{\xb}{ \bar{x} }
\newcommand{\bb}{ \bar{b} }
\newcommand{\vphib}{ \bar{\varphi} }
\newcommand{\Phib}{ \bar{\Phi} }

\newcommand{\qCG}{ q_{\scriptscriptstyle CG} }

\usepackage{xcolor}

\usepackage{cancel}
\usepackage[normalem]{ulem}

\subheader{\hfill USTC-ICTS-15-01}

\title{\boldmath Phase structures of the black \Dp-\Dppf-brane system in various ensembles I:
thermal stability}

\author[a,b]{Da Zhou}
\author[a,c]{and Zhi-Guang Xiao}

\affiliation[a]{The Interdisciplinary Center for Theoretical Study,\\
University of Science and Technology of China,\\
Hefei, Anhui, 230026, China}

\affiliation[b]{Department of Mathematics, City University London,\\
London EC1V 0HB, U.K.}

\affiliation[c]{State Key Laboratory of Theoretical Physics,\\
Institute of Theoretical Physics,\\
Chinese Academy of Sciences, Beijing 100190, China}
  
\emailAdd{da.z.zhou@gmail.com}
\emailAdd{xiaozg@ustc.edu.cn}

\abstract{When the D$(p+4)$-brane ($p=0,1,2$) with delocalized D$p$ charges  is put into
equilibrium with a spherical thermal cavity, the two kinds of charges
can be put into canonical or grand canonical ensemble independently by
setting different conditions at the boundary. Using the thermal stability
condition, we discuss the
phase structures of various ensembles of this system formed in this
way and find out the situations that the black brane could be the
final stable phase in these ensembles.  In particular, van der
Waals-like phase transitions can happen when D0 and D4 charges are in
different kinds of ensembles.  Furthermore, our results indicate that
the \Dppf-branes and the delocalized \Dp-branes are equipotent. 
}

\begin{document}

\maketitle

\section{Introduction}
Branes are non-perturbative objects in String theory and study on
them is helpful in 
understanding  the non-perturbative properties of String theory.
Various brane configurations are also useful in the applications of
gauge-string duality. Under the assumption of AdS/CFT, study on the
near horizon geometry of D3 brane, i.e. AdS$_5\times S^5$, reveals the
properties of corresponding strong coupling conformal field theory in
the large $N_c$ limit at the boundary of AdS$_5$, which can also be
extended to the asymptotically AdS$_5$ spaces. A typical example is
that the well-known Hawking-Page phase transition of the AdS black
hole corresponds to the confinement-deconfinement phase transition in
$SU(N_c)$ gauge theory at large $N_c$ \cite{Witten:1998zw}, which
aroused the recent interests on the black hole thermodynamics in the
asymptotically AdS space.

As the solutions of the  supergravity, black branes, like black holes,
can have their own thermodynamics.  Study on the thermodynamical phase
structure of the black branes is also valuable in understanding the
non-perturbative nature of String theory.  
Since black branes are asymptotically flat,
they are unstable by themselves due to their negative specific heat capacity
and the Hawking radiation. To study their phase structures, one can put
them into a spherical cavity which is considered as a reservoir
to form a thermal equilibrium, following the approach of York in discussing the
thermodynamics of black
holes\cite{York:1986it,Whiting:1988qr,Braden:1990hw}. For different
boundary conditions at the boundary of the cavity, there can be
different ensembles  
for charged black
branes.  Along these lines, in \cite{lu:2011}, the phase structure of
black $p$-branes in the canonical ensemble was studied in
$D$-dimensional space-time, where the temperature, the volume of the
brane as well as the cavity, and the charges of the brane are fixed.
There can be van der Waals-like phase transition in this system for
$\tilde d>2$ ($p<5$) where $D=d+\tilde d+2$ and $ p=d-1$: a first order phase
transition between a large black brane and a small one can be found
for charge $q<q_c$, while for $q>q_c$ there is only one black
brane phase, with $q_c$ being the critical charge at which there is a
second order phase transition.  This is similar to the van der Waals-like phase
transition found in charged AdS black holes
\cite{Chamblin:1999tk,Chamblin:1999hg}, asymptotically flat
black holes as well as dS black
holes \cite{Carlip:2003ne,Lundgren:2006kt}.  In contrast, for $\tilde d \leq 2$
($p=5,6$),
there is no van der Waals-like phase transition. For uncharged black
branes, Hawking-Page like
transitions between black branes and the ``hot flat space'' can also
happen in the canonical ensemble similar to the uncharged black
hole case \cite{York:1986it,Whiting:1988qr}. The phase structure for
black branes in grand canonical ensemble is different from that in  canonical ensemble for $\tilde d >2$, where the
potential is fixed while the charges are not. There is no van der
Waals-like
phase transition, but the Hawking-Page-like transitions between the
black branes and the ``hot flat space'' can happen. Bubble
solutions\cite{Witten:1981gj,Horowitz:2005vp} which can be obtained
from black branes by double wick rotations also play a role in the
phase structure of black branes, since they have the same boundary
condition as the black branes. In \cite{Lu:2011da} and
\cite{Wu:2011yu}, bubbles were found to be the phases of black branes in
canonical ensembles and grand canonical ensembles, and can have phase
transitions with black branes, and thus enriches the phase structure of
 black branes. In all these discussions, 
there are no van der Waals-like phase transitions in  grand
canonical ensembles, whereas they do exist in canonical ensembles. The
absence of van der Waals-like phase transitions in grand canonical ensemble
is also true for charged black holes either in flat
space\cite{Braden:1990hw,Carlip:2003ne} or in AdS space
\cite{Chamblin:1999tk,Chamblin:1999hg}, but they may exist for
Gauss-Bonnet AdS black
holes\cite{Dey:2007vt,Anninos:2008sj,Zou:2014mha} for certain
Gauss-Bonnet coupling constants.

It is well-known that branes can be combined to form composite states
or intersecting branes. Among others, D$p$-D$(p+4)$ bound states as
solutions of supergravities can be constructed by smearing D$p$ branes
inside D$(p+4)$ branes. In extremal
cases\cite{Callan:1996dv,Behrndt:1996pm,Papadopoulos:1996ca} which
obey the harmonic function rules,
the solutions preserve $1/4$ of the supersymmetries. There can also be
more generic non-supersymmetric D$p$-D$(p+4)$ solutions in ten-dimensional
supergravities\cite{Costa:1996re,Peet:2000hn,Miao:2004bn}. These brane solutions are useful in
gravity-gauge duality
applications\cite{Liu:1999fc,Barbon:1999zp,Suzuki:2000sv,Wu:2013zxa,Cai:2014wia,Seki:2013nta} and in constructing lower
dimensional black holes\cite{Peet:2000hn}.  The dynamical stability of D$p$-D$(p+4)$ was discussed
in\cite{Friess:2005tz}. We are interested in the thermodynamical phase
structure of these  solutions in this paper. Since these solutions
involve two kinds of charges coupled with different RR potentials, the
phase structure is expected to be richer. One  can form different
ensembles by fixing either the charge inside the cavity or the
potential at the boundary for D$p$ or/and D$(p+4)$ branes , and therefore
D$p$ and D$(p+4)$ can be in canonical ensemble or in grand canonical
ensemble independently.  In \cite{lu:2012-2}, the phase structure of
the D1-D5 system in which both branes are in the canonical ensemble is
discussed.  It has been shown that the smeared D1-brane alone shares the
same phase structure as the D5-brane for which the van der Waals-like phase
structure can not be found.  However the D1-D5 combined system
displays a van der Waals-like phase structure for some region of the charge
combinations. And the phase structure is also symmetric under the exchange
of the D1 charges and D5 charges. One may also be curious about what
will happen when the two kinds of charges are in different ensembles.
In present paper we will study the phase structures of the other three
combinations of the ensembles: both D$p$ and D$(p+4)$ in grand canonical
ensemble, D$p$ in canonical and D$(p+4)$ in grand canonical
ensemble, and D$p$ in grand canonical
and D$(p+4)$ in canonical ensemble.   We will find that the phase structures
are also symmetric when one swaps the boundary conditions imposed on D$p$- and
D$(p+4)$-branes. For mixed ensembles with one charge in canonical
ensemble and the other in grand canonical ensemble, one may expect in
some charge combinations there could be van der Waals-like phase
transitions like the canonical ensemble  and in some other regions
there are no van der Waals-like phase transitions as in grand canonical
ensemble. This indeed happens in the D0-D4 system, which has the
richest phase structure. In D2-D6 and D1-D5 systems, unlike in the D1-D5
canonical ensemble, in the other three ensembles there are still no
van der Waals-like phase transitions. In our discussion, we only
consider the stability condition of the brane under a small change of
the horizon size, which is equivalent to the positive specific heat capacity
condition. Nonetheless, we will briefly comment on the effect of the more
generic stability conditions --- the electrical stability conditions for
both charges --- on the phase structure. In the meantime, the technical details
on this electrical stability conditions is arranged in another paper \cite{xiao:2015}.

The paper is organized as follows. In section \ref{sect:Branes}, we
review the D$p$-D$(p+4)$ solution and evaluate the classical Euclidean actions or
 thermodynamic potentials for different ensembles. In section
\ref{sect:Ensembles}, we discuss the phase structures of the
D$p$-D$(p+4)$ system in different ensembles.
In section \ref{sect:General-stability}, we give a brief discussion on
more general thermodynamic stability conditions.
Section \ref{sect:Conclude} is the conclusion. We also gather some
detailed calculations in the Appendices.

\section{The D$p$-D$(p+4)$ brane system\label{sect:Branes}}

\subsection{The action}

We consider a gravitational system bounded by a big spherical reservoir
which can be regarded as a spherical boundary at the transverse radius
$\rho_b$. At the center of the system is a pile of parallel
\Dp-\Dppf-branes. The
total Euclidean 10-dimensional supergravity action in Einstein frame can be expressed as a sum of several
contributions~\cite{lu:2011},
\begin{equation}
  I = \IEH + I_\phi + I_p + I_{p+4} + \IB
  \label{eq:total-action}
\end{equation}
where the first term is the usual Einstein-Hilbert action,
\begin{equation}
  \IEH = - \frac{1}{2\kappa^2} \intx \sqrt{g} R ,
  \label{eq:einstein-hilbert}
\end{equation}
the second term is  the contribution from the dilaton field,
\begin{equation}
  I_\phi = \frac{1}{4\kappa^2} \intx \sqrt{g} \partial^\mu\phi \partial_\mu\phi ,
  \label{eq:dilaton}
\end{equation}
the third and fourth terms come from the \Dp- and \Dppf-brane form field action,
\begin{eqnarray}
  I_p & = & \frac{1}{4\kappa^2} \intx \sqrt{g} \, 
  \frac{e^{a_p\phi}}{(p+2)!} F_{[p+2]}^2 ,\nonumber\\
  I_{p+4} & = & \frac{1}{4\kappa^2} \intx \sqrt{g}
  \frac{e^{a_{p+4}\phi}}{(p+6)!} F_{[p+6]}^2 ,
  \label{eq:gauge-fields-action}
\end{eqnarray}
where $a_n=\frac{3-n}{2}$, and the last term $\IB$ could admit several boundary
integration terms, which depends on what ensemble we are interested in.
We will come back to this term soon. In the above formulae, the
constant coefficient $\kappa$ is defined as $\kappa=\sqrt{8\pi
G_{10}^2}$ in which $G_{10}$ is the 10-dimensional Newton's constant;
the vacuum expectation value of the dilaton field 
$\phi(r\!\rightarrow\!\infty)$ has been shifted to zero. The integration
is performed within the boundary and outside the horizon if there exists
one. 

Now we deal with the boundary term in equation (\ref{eq:total-action}).
There are three terms that may contribute to the boundary action $\IB$,
\begin{eqnarray}
  \IGH & = & \frac{1}{\kappa^2} \ointx \sqrt{\gamma} \, (K-K_0) ,\nonumber\\
  I_{b,p} & = & - \frac{1}{2\kappa^2} \ointx \sqrt{\gamma} \,
  \frac{e^{a_p\phi}}{(p+1)!} n_\mu F^{\mu\nu_1\cdots\nu_{p+1}}
  A_{\nu_1\cdots\nu_{p+1}} ,\nonumber\\
  I_{b,p+4} & = & - \frac{1}{2\kappa^2} \ointx \sqrt{\gamma}
  \frac{e^{a_{p+4}\phi}}{(p+5)!} n_\mu F^{\mu\nu_1\cdots\nu_{p+5}}
  A_{\nu_1\cdots\nu_{p+5}} .
  \label{eq:boundary-terms}
\end{eqnarray}
The first action above is the Gibbons-Hawking surface term~\cite{gibbons:1977},
in which $K$ is the trace of the extrinsic curvature $K_{\mu\nu}$ defined as
\begin{equation}
K_{\mu\nu} = -\frac{1}{2} ( \nabla_\mu n_\nu + \nabla_\nu n_\mu )
\end{equation}
where $n_\mu$ is the normalized space-like vector normal to the boundary.
$K_0$ in that term is defined in the same manner as $K$ but with the
metric replaced with flat metric. This subtraction term is included
to make $\IGH$ vanish for flat metric~\cite{gibbons:1977}.  The second
and third actions in (\ref{eq:boundary-terms}) are boundary contributions
from the \Dp- and \Dppf-branes respectively. $\gamma$ in these equations
is the determinant of the induced metric on the $(4-p)$-dimensional boundary,
and the $(p+1)$-form and $(p+5)$-form fields $A_{[p+1]}$ and $A_{[p+5]}$
are the Ramond-Ramond potentials of \Dp- and \Dppf-branes respectively.

The Gibbons-Hawking term should always be included in $\IB$ regardless of
what ensemble we are talking about. However, $I_{b,p}$ and $I_{b,p+4}$
are supposed to cancel additional boundary terms when doing variations
with respect to the gauge field potentials.
If we fix the gauge field strength $F$ on the boundary and the potential $A$ are not fixed,
 after partial integration the additional boundary terms would emerge,
thus we need to include $I_{b,p}$ or $I_{b,p+4}$ whose variations would cancel these terms.
On the other hand, if we fix the gauge potential $A$ on the boundary,
there will be no additional variation terms, therefore,
$I_{b,p}$ or $I_{b,p+4}$ will not be needed.
We summarize the above analyses in Table \ref{tb:bc-ba-relation}.
\begin{table}[!ht]
\begin{center}
  \begin{tabular}{c|c|c}
  \hline
  \Dp & \Dppf & Boundary action $\IB$ \\
  \hline
  Fixing $A_{[p+1]}$ & Fixing $A_{[p+5]}$ & $\IGH$ \\
  \hline
  Fixing $F_{[p+2]}$ & Fixing $A_{[p+5]}$ & $\IGH+I_{b,p}$ \\
  \hline
  Fixing $A_{[p+1]}$ & Fixing $F_{[p+6]}$ & $\IGH+I_{b,p+4}$ \\
  \hline
  Fixing $F_{[p+2]}$ & Fixing $F_{[p+6]}$ & $\IGH+I_{b,p}+I_{b,p+4}$ \\
  \hline
  \end{tabular}
  \caption{The relationship between boundary conditions and the boundary action \label{tb:bc-ba-relation}}
\end{center}
\end{table}
The relation between boundary conditions and ensembles will be addressed in next subsection.

\subsection{Black brane solution}

The generic non-supersymmetric D$p$-D$(p+4)$ brane solution
\cite{Costa:1996re,Peet:2000hn,Miao:2004bn,Friess:2005tz} can be obtained by directly solving the
supergravity equations of motion or by a series of duality and boost
operations from D$(p+4)$-branes.
We use following coordinates
to describe the solution,
\begin{equation}
  (t, x_1, \cdots, x_p, \cdots, x_{p+4}, \rho, \phi_1, \cdots, \phi_{4-p})
  \label{eq:coordinates}
\end{equation}
where $\rho$ is the radius in the transverse directions perpendicular
to $x_m$, $m=1,\cdots,p+4$.  In this coordinates the solution reads
\cite{lu:2012-2},
\begin{eqnarray}
  ds^2 & = & \DM^{\frac{1-p}{4}} \DS^{\frac{p-7}{8}}
  \left( \DP dt^2 + \DM \sum^{p}_{i=1} dx_i^2
  + \DS \sum^{p+4}_{j=p+1} dx_j^2 \right) \nonumber\\
  & & + \DM^{\frac{p^2-1}{4(3-p)}} \DS^{\frac{p+1}{8}}
  \left( \frac{d\rho^2}{\DP\DM} + \rho^2 d\Omega_{4-p}^2 \right) ,\nonumber\\
  e^\phi &=& \DM^{\frac{p-1}{2}} \DS^{\frac{3-p}{4}} ,\nonumber\\
  A_{[p+1]} &=& -i \frac{\DP}{\DS} \left( \frac{\DS-\DM}{\DS-\DP} \right)^{1/2}
  dt \wedge dx_1 \wedge \cdots \wedge dx_p ,\nonumber\\
  F_{[p+2]} &=& i \frac{3-p}{\rho} \frac{1}{\DS}
  \left( 1- \frac{\DP}{\DS} \right)^{1/2}
  \left( 1- \frac{\DM}{\DS} \right)^{1/2}
  dt \wedge d\rho \wedge dx_1 \wedge \cdots \wedge dx_p ,\nonumber\\
  A_{[p+5]} &=& -i \DP \left( \frac{1-\DM}{1-\DP} \right)^{1/2}
  dt \wedge dx_1 \wedge \cdots \wedge dx_{p+4} ,\nonumber\\
  F_{[p+6]} & = & i \frac{3-p}{\rho} (1-\DP)^{1/2} (1-\DM)^{1/2}
  \ dt \wedge d\rho \wedge dx_1 \wedge \cdots \wedge dx_{p+4} ,
  \label{eq:solution}
\end{eqnarray}
where
\begin{eqnarray}
  \Delta_\pm(\rho) &=&  1 - \frac{\rho_\pm^{3-p}}{\rho^{3-p}} ,
  \qquad \RP > \RM \geq 0 ,\nonumber\\
  \DS(\rho) &=&  1 - \frac{k}{\rho^{3-p}} ,
  \qquad \RM^{3-p} \geq k > -\infty .
  \label{eq:deltas}
\end{eqnarray}
Here, $p$ can be $0,1,2$ for this  
supergravity solution to describe the D$p$-D$(p+4)$ system. The constants $\RP$ and $\RM$ are the coordinates for the outer event
horizon and an inner event horizon respectively,  the latter being a
curvature singularity.  Therefore, the requirement $\RP>\RM$ is to
avoid a naked curvature
singularity. The other constant $k$ can be either positive or negative, and
when it is positive, there is another inner event horizon at
$\rho=k^{1/(3-p)}$ which is a curvature singularity as well.  When
$k=\RM^{3-p}$, we have $\DS=\DM$, and this corresponds to the brane
configuration where all \Dp-branes are removed from the system while
\Dppf-branes are retained. Later we will see in (\ref{eq:charge-density})
that we require $\RM^{3-P}\ge k$ in order to make the Ramond-Ramond charge real.

The \Dp- and \Dppf-brane charge densities can be obtained by integrating
the Hodge dual of their Ramond-Ramond field strengths,
\begin{eqnarray}
  \Q_p &=&  \frac{-i}{\sqrt{2}\kappa} \int_{x_{p+1},\cdots,x_{p+4}} \oint_{S^{4-p}}
  e^{a_p\bphi} *F_{[p+2]}(\rho_b) ,\nonumber\\
  \Q_{p+4} &=& \frac{-i}{\sqrt{2}\kappa} \oint_{S^{4-p}}
  e^{a_{p+4}\bphi} *F_{[p+6]}(\rho_b) ,
  \label{eq:charge-formulae}
\end{eqnarray}
where $\bphi\equiv\phi(\rho_b)$ and the $-i$ factors come from Euclideanization.
The charge densities can be explicitly expressed as follows,
\begin{eqnarray}
  \Q_p &=& \frac{(3-p)V_4\Omega_{4-p}}{\sqrt{2}\kappa} e^{a_p\bphi/2} \brho_b^{3-p}
  \left( 1- \frac{\BDP}{\BDS} \right)^{1/2} \left( 1- \frac{\BDM}{\BDS} \right)^{1/2} ,\nonumber\\
  \Q_{p+4} &=& \frac{(3-p)\Omega_{4-p}}{\sqrt{2}\kappa} e^{a_{p+4}\bphi/2} (\BRP\BRM)^{\frac{3-p}{2}} ,
  \label{eq:charge-density}
\end{eqnarray}
where
\begin{equation}
  V_4 = \sqrt{g_{x_{p+1}x_{p+1}}(\rho_b) \cdots g_{x_{p+4}x_{p+4}}(\rho_b)}\ V_4^*
  ,\qquad V_4^* \equiv \int dx_{p+1} \cdots dx_{p+4} ,\nonumber\\
\end{equation}
and
\begin{eqnarray}
  \bar{\Delta}_\pm &\equiv& \Delta_\pm(\rho_b) = 1 - \frac{\brho_\pm^{3-p}}{\brho_b^{3-p}} ,\nonumber\\
  \BDS &\equiv& \DS(\rho_b) = 1 - \frac{\bar{k}}{\brho_b^{3-p}} ,
  \label{eq:def-bar}
\end{eqnarray}
in which
\begin{equation}
  \brho_{+,-,b} \equiv \rho_{+,-,b} \, \BDM^{\frac{p^2-1}{8(3-p)}}
  \BDS^{\frac{p+1}{16}} = \rho_{+,-,b} \, e^{\frac{p+1}{4(3-p)}\bphi}
  \label{eq:def-bar-rho}
\end{equation}
are the physical radii of horizons and the boundary, and
$\bar{k} \equiv k\, e^{\frac{p+1}{4}\bphi}$.

To make these relations more concise, we define two reduced charge densities,
\begin{eqnarray}
  \TQ_p &=& \frac{\sqrt{2}\kappa\Q_p e^{-a_p\bphi/2}}{(3-p)
    V_4 \Omega_{4-p} \brho_b^{3-p}} \nonumber\\
  &=& \left( 1- \frac{\BDP}{\BDS} \right)^{1/2}
  \left( 1- \frac{\BDM}{\BDS} \right)^{1/2} < 1 ,\nonumber\\
  \TQ_{p+4} &=& \left( \frac{\sqrt{2}\kappa\Q_{p+4}
  e^{-a_{p+4}\bphi/2}}{(3-p)\Omega_{4-p}} \right)^{\frac{1}{3-p}}
  = \sqrt{\BRP\BRM} .
  \label{eq:reduced-charge-density}
\end{eqnarray}
With (\ref{eq:reduced-charge-density}) we can express $\BRM$ and
$\BDS$ through $\BRP$ and these two reduced quantities,
\begin{eqnarray}
  \BRM &=& \frac{\TQ_{p+4}^2}{\BRP} ,\nonumber\\
  \BDS &=& \frac{\BDP + \BDM \pm \sqrt{(\BDM-\BDP)^2 + 4\TQ_p^2 \BDP\BDM}}{2(1-\TQ_p^2)} .
  \label{eq:brm-bds}
\end{eqnarray}
When $k=\RM^{3-p}$, the \Dp-brane charge vanishes and $\TQ_p=0$, so
\[ \BDS = \frac{\BDP + \BDM \pm ( \BDM - \BDP )}{2} = \BDM . \]
Therefore, we should choose in the above relation the ``+'' sign
only, i.e.,
\begin{eqnarray}
  \BDS &=& \frac{\BDP + \BDM + \sqrt{(\BDM-\BDP)^2 + 4\TQ_p^2 \BDP\BDM}}{2(1-\TQ_p^2)} .
  \label{eq:delta-star}
\end{eqnarray}

From (\ref{eq:charge-density}), we can see that fixing $F$ at the boundary
is to fix the charge density $\Q$ within the boundary.  It is well-known
that fixing the charge of a system while allowing radiation means we are
considering the system in canonical ensemble. If we fix the gauge potential
$A$ instead, the field strength at the boundary will be changeable, and
we will be talking about grand canonical ensemble. Now that we have two kinds
of gauge potentials and field strengths, it is reasonable to assume there
are two kinds of ensembles for each field, and thus we have the following table
(Table \ref{tb:ens-ba-relation}).
\begin{table}[t]
\begin{center}
  \begin{tabular}{c|c|c}
  \hline
  \Dp-brane & \Dppf-brane & Boundary action $\IB$ \\
  \hline
  Canonical & Canonical & $\IGH+I_{b,p}+I_{b,p+4}$ \\
  \hline
  Grand Canonical & Canonical & $\IGH+I_{b,p+4}$ \\
  \hline
  Canonical & Grand Canonical & $\IGH+I_{b,p}$ \\
  \hline
  Grand Canonical & Grand Canonical & $\IGH$ \\
  \hline
  \end{tabular}
  \caption{The relationship between ensembles and the boundary action \label{tb:ens-ba-relation}}
\end{center}
\end{table}

\subsection{Temperature and conjugate potentials}

Like black holes,  due to Hawking radiation \cite{hawking:1975},  black brane
has a temperature, $T_H$, which can be easily calculated by the
requirement that  the Euclideanized metric (\ref{eq:solution}) has no
conical singularity. Then the Euclideanized time $t$ is cyclic with a
particular period
\begin{equation}
  \beta^* = \frac{4\pi \rho_b}{3-p} \frac{(1-\BDP)^{1/2}(\BDS-\BDP)^{1/2}}{(\BDM-\BDP)^{\frac{2-p}{3-p}}} .
  \label{eq:time-period}
\end{equation}
This period 
 is just the inverse of the  Hawking temperature observed by an observer at $\rho=\infty$,
\begin{equation}
  T_H = 1 / \beta^* .
  \label{eq:hawking-temperature}
\end{equation}
 For a local
observer at $\rho=\rho_b$, the local temperature would be
\begin{equation}
  \bbeta = \beta^* \BDP^{\frac{1}{2}} \BDM^{\frac{1-p}{8}} \BDS^{\frac{p-7}{16}} =
  \frac{4\pi\brho_b}{3-p} (1-\BDP)^{\frac{1}{2}} \left( \frac{\BDP}{\BDM} \right)^{\frac{1}{2}}
  \left( 1- \frac{\BDP}{\BDM} \right)^{\frac{p-2}{3-p}} \left( 1- \frac{\BDP}{\BDS} \right)^{\frac{1}{2}} .
  \label{eq:local-temperature}
\end{equation}

In order to study the grand canonical ensembles, we need to define two
potentials which are conjugate to the brane charge densities. So we
define $\Phi_p$  to be the potential conjugate to the D$p$ charge 
 using 
\begin{equation}
  A_{[n+1]} = -i \sqrt{2} \kappa \,\Phi_n\, d\bar{t} \wedge d\bar{x}_1 \wedge \cdots \wedge d\bar{x}_n ,
  \label{eq:conjugate-potential}
\end{equation}
where the barred coordinates are defined as follows,
\begin{eqnarray}
  \bar{t} &=&  t \, \BDP^{\frac{1}{2}} \BDM^{\frac{1-p}{8}} \BDS^{\frac{p-7}{16}} ,\nonumber\\
  \bar{x}_i &=&  x_i \, \BDM^{\frac{5-p}{8}} \BDS^{\frac{p-7}{16}} ,\qquad i=1,\cdots,p ,\nonumber\\
  \bar{x}_j &=&  x_j \, \BDM^{\frac{1-p}{8}} \BDS^{\frac{p+1}{16}} ,\qquad j=p+1,\cdots,p+4 .
  \label{eq:barred-coordinates}
\end{eqnarray}
The reasonableness of the definition of the conjugate potentials will be
justified later. At the boundary, in the equilibrium state, $\Phi_p$
and $\Phi_{p+4}$ can be expressed using $\bar \Delta_{+,-,*}$   
explicitly according to (\ref{eq:solution}),
\begin{eqnarray}
  \Phi_p &=& \frac{1}{\sqrt{2}\kappa} e^{-a_p\bphi/2}
  \left( \frac{\BDP}{\BDM} \right)^{1/2} \left( \frac{\BDS-\BDM}{\BDS-\BDP} \right)^{1/2} ,\nonumber\\
  \Phi_{p+4} &=& \frac{1}{\sqrt{2}\kappa} e^{-a_{p+4}\bphi/2}
  \left( \frac{\BDP}{\BDM} \right)^{1/2} \left( \frac{1-\BDM}{1-\BDP} \right)^{1/2} . 
  \label{eq:phi-explicit}
\end{eqnarray}

\subsection{Evaluation of actions}

Since the bulk action terms and the Gibbons-Hawking term in the total
action (\ref{eq:total-action}) are the common parts that appear in
every ensemble, we define their sum as
\begin{equation}
  I_c = I_{\rm EH} + I_\phi + I_p + I_{p+4} + I_{\rm GH}
  \label{eq:common-action}
\end{equation}
for later convenience. Using the solution (\ref{eq:solution}) one can
evaluate the actions, 
\begin{eqnarray}
  I_c & = & - \frac{\bbeta V_{p+4} \Omega_{4-p}}{2\kappa^2} \brho_b^{3-p}
  \left[ (5-p) \left( \frac{\BDP}{\BDM} \right)^{\frac{1}{2}} +
    (3-p) \left( \frac{\BDM}{\BDP} \right)^{\frac{1}{2}} - 2(4-p) \right] \nonumber\\
  &=& - \frac{\bbeta V_{p+4} \Omega_{4-p}}{\kappa^2} \brho_b^{3-p}
    \left[ (4-p) \left( \frac{\BDP}{\BDM} \right)^{\frac{1}{2}} - (4-p) \right] - S ,\nonumber\\
  I_{b,p} & = & - \frac{\bbeta V_{p+4} \Omega_{4-p}}{2\kappa^2} \brho_b^{3-p}
    \left[ (3-p) \frac{(\BDP\BDM)^{\frac{1}{2}}}{\BDS} - (3-p) \left( \frac{\BDP}{\BDM} \right)^{\frac{1}{2}} \right] 
    = \bbeta V_p \Q_p \Phi_p , \nonumber\\
  I_{b,p+4} & = & - \frac{\bbeta V_{p+4} \Omega_{4-p}}{2\kappa^2} \brho_b^{3-p}
    \left[ (3-p) (\BDP\BDM)^{\frac{1}{2}} - (3-p) \left( \frac{\BDP}{\BDM} \right)^{\frac{1}{2}} \right]
    = \bbeta V_{p+4} \Q_{p+4} \Phi_{p+4} ,
\end{eqnarray}
where
\begin{equation}
  S = \frac{2\pi V_{p+4} \Omega_{4-p}}{\kappa^2} \brho_b^{4-p} (1-\BDP)^{\frac{1}{2}}
  \left( 1- \frac{\BDP}{\BDM} \right)^{\frac{1}{3-p}} \left( 1- \frac{\BDP}{\BDS} \right)^{\frac{1}{2}}
\end{equation}
is the entropy of the brane system and
\begin{equation}
  V_{p+4} = V_p V_4 ,\qquad V_p = \sqrt{g_{x_1x_1}(\rho_b) \cdots g_{x_px_p}(\rho_b)}\ V_p^* ,\qquad V_p^* = \int dx_1 \cdots dx_p .
\end{equation}
With the above results and Table~\ref{tb:ens-ba-relation}, we obtain the
 classical actions for various ensembles:
\begin{itemize}
  \item Both charges are in canonical ensembles (we will use CC ensemble to
denote this one)
\begin{eqnarray}
  \ICC & = & I_c + I_{b,p} + I_{b,p+4} \nonumber\\
  & = & - \frac{\bbeta V_{p+4} \Omega_{4-p}}{2\kappa^2} \brho_b^{3-p}
  \left[ 2\left( \frac{\BDP}{\BDM} \right)^{\frac{1}{2}}
  + (3-p) (\BDP\BDM)^{\frac{1}{2}} \left( 1+ \frac{1}{\BDS} \right) +2p - 8 \right] -S \nonumber\\
  &=& \bbeta E - S ,
  \label{eq:canon-canon}
\end{eqnarray}
where
\begin{equation}
  E = - \frac{V_{p+4} \Omega_{4-p}}{2\kappa^2} \brho_b^{3-p} \left[ 2\left( \frac{\BDP}{\BDM} \right)^{\frac{1}{2}}
  + (3-p) (\BDP\BDM)^{\frac{1}{2}} \left( 1+ \frac{1}{\BDS} \right) +2p - 8 \right]
  \label{eq:internal-energy}
\end{equation}
is the internal energy of the brane system. This action has already been
obtained in \cite{lu:2012-2}, and we include it here for completeness.

\item \Dp\ in grand canonical ensemble \& \Dppf\ in canonical ensemble
(denoted as GC ensemble )
\begin{equation}
  \IGC 
= \bbeta E - S - \bbeta V_p \Q_p \bar \Phi_p .
  \label{eq:grand-canon}
\end{equation}

\item \Dp\ in canonical ensemble \& \Dppf\  in
grand canonical ensemble (denoted as CG
ensemble)
\begin{equation}
  \ICG
= \bbeta E - S - \bbeta V_{p+4} \Q_{p+4} \bar\Phi_{p+4} .
  \label{eq:canon-grand}
\end{equation}

\item Both charges are in grand canonical ensemble (denoted as GG
ensemble)
\begin{equation}
  \IGG
= \bbeta E - S - \bbeta V_p \Q_p \bar\Phi_p - \bbeta
V_{p+4} \Q_{p+4} \bar\Phi_{p+4} .
  \label{eq:grand-grand}
\end{equation}
\end{itemize}
Here, $\bar\Phi_p$ and $\bar\Phi_{p+4}$ are the potentials imposed on
the boundary and we stick to the convention that the barred quantities
are the ones on the boundary. 
Only at equilibrium,  $\bar\Phi_p=\Phi_p$ and
$\bar\Phi_{p+4}=\Phi_{p+4}$, and 
\begin{eqnarray}
\IGC = I_c + I_{b,p+4} ,\quad \ICG = I_c + I_{b,p}\, ,\quad \IGG =
I_c.
\end{eqnarray}
It is easy to check that, for $\rho_b\rightarrow\infty$,
(\ref{eq:internal-energy}) reduces to
\begin{equation}\label{eq:adm-mass}
  E \eval{\rho_b\rightarrow\infty} = \frac{V_p^* V_4^* \Omega_{4-p}}{2\kappa^2}
  \Big[ (4-p) \RP^{3-p} + (2-p) \RM^{3-p} - (3-p)k \Big]
\end{equation}
the right hand side of which is exactly the ADM mass of the branes.
We know that  the free energy equals to
its internal energy minus the temperature times the
entropy, where the internal energy is the ADM mass for a gravitational
system according to Bardeen et. al. \cite{bardeen:1973}. So (\ref{eq:canon-canon}) and
(\ref{eq:adm-mass}) means
that we are indeed dealing with a system in canonical ensemble.

Now we justify that (\ref{eq:canon-canon}) (\ref{eq:grand-canon})
(\ref{eq:canon-grand}) and (\ref{eq:grand-grand}) are indeed the
correct forms of free energy or grand potential, and the conjugate
potentials defined in (\ref{eq:conjugate-potential}) and
(\ref{eq:phi-explicit}) are consistent. For this purpose, we will
check that the equilibrium corresponds to the stationary point in
various ensembles. First we take the
derivative of $\ICC$ with respect to $\BRP$ and set it to zero,
\begin{equation}
  \PP{\ICC}{\BRP} = \frac{2\pi(3-p)V_{p+4}\Omega_{4-p}\BRP^{2-p}\brho_b}{\kappa^2} 
  f\big(\BRP,\TQ_p,\TQ_{p+4}\big) \Big[ \bar{b} - b\big(\BRP,\TQ_p,\TQ_{p+4}\big) \Big] = 0 ,
  \label{eq:dICC-dBRP}
\end{equation}
where
\begin{eqnarray}
  \bar{b} &=& \frac{\bbeta}{4\pi\brho_b} ,\nonumber\\
  b\big(\BRP,\TQ_p,\TQ_{p+4}\big) &=& \frac{1}{3-p} (1-\BDP)^{\frac{1}{2}} \left( \frac{\BDP}{\BDM} \right)^{\frac{1}{2}}
  \left( 1- \frac{\BDP}{\BDM} \right)^{\frac{p-2}{3-p}} \left( 1- \frac{\BDP}{\BDS} \right)^{\frac{1}{2}} ,\nonumber\\
  f\big(\BRP,\TQ_p,\TQ_{p+4}\big) &=& \frac{1}{2 \BDP^{1/2} \BDM^{3/2} (1-\BDP)}
  \Big[ (3-p) \BDM (\BDM-\BDP) + 2(\BDP+\BDM-2\BDP\BDM) \nonumber\\
  && + (3-p) \frac{\BDM(\BDM-\BDP)(\BDP+\BDM-2\BDP\BDM)}{\BDS\BDP+\BDS\BDM-2\BDP\BDM} \Big] .
  \label{eq:def-b-f}
\end{eqnarray}
The solution to (\ref{eq:dICC-dBRP}) is the stationary point of the
action, which is supposed to correspond to the equilibrium
state, and is thus expected to give the equation of state
(\ref{eq:local-temperature}). Since $\BDP<\BDM\leq\BDS\leq 1$, which can
be seen from (\ref{eq:deltas}), the function $f$ defined in the
last equation of (\ref{eq:def-b-f}) is positive definite. This
means the expression in the square brackets of (\ref{eq:dICC-dBRP})
must vanish, which recovers the equation of state
(\ref{eq:local-temperature}) exactly. This justifies our claim
that $\ICC$ is the correct form of free energy.

Next we take the derivative of $\IGC$ with respect to $\Q_p$ and
again set it to zero,
\begin{eqnarray}
  \PP{\IGC}{\Q_p} &=& - \frac{\bbeta V_p}{\sqrt{2}\kappa e^{a_p\bphi/2}}
  \Bigg\{ \bar{\Phi} - \Phi\big(\BRP,\TQ_p,\TQ_{p+4}\big)
  \Bigg[ 1 + \nonumber\\
  && \Bigg( 1 - \frac{b\big(\BRP,\TQ_p,\TQ_{p+4}\big)}{\bar{b}} \Bigg)
  \frac{\BDS(\BDM-\BDP)}{\BDS\BDP+\BDS\BDM-2\BDP\BDM} \Bigg] \Bigg\} \nonumber\\
  &=& - \frac{\bbeta V_p}{\sqrt{2}\kappa e^{a_p\bphi/2}}
  \left[ \bar{\Phi} - \Phi\big(\BRP,\TQ_p,\TQ_{p+4}\big) \right] \nonumber\\
  &=& 0
  \label{eq:dIGC-dQp}
\end{eqnarray}
where
\begin{eqnarray}
  \bar{\Phi} &=& \sqrt{2} \kappa e^{a_p \bphi/2} \bar\Phi_p ,\nonumber\\
  \Phi\big(\BRP,\TQ_p,\TQ_{p+4}\big) &=& \left( \frac{\BDP}{\BDM} \right)^{1/2} \left( \frac{\BDS-\BDM}{\BDS-\BDP} \right)^{1/2} .
  \label{eq:def-Phi}
\end{eqnarray}
In getting the first equality in (\ref{eq:dIGC-dQp}), we have used the expression for
$\TQ_p$ in (\ref{eq:reduced-charge-density}), and in the
second equality, we have used the fact that
$\bar{b}=b(\BRP,\TQ_p,\TQ_{p+4})$ in equilibrium state.  The last
equality in (\ref{eq:dIGC-dQp}) just recovers the first equation
in (\ref{eq:phi-explicit}) where it is merely a definition. This
proves the validity of that definition. Similarly, we
calculate the derivative of $\ICG$ with respect to $\Q_{p+4}$,
\begin{eqnarray}
  \PP{\ICG}{\Q_{p+4}} &=& - \frac{\bbeta V_p}{\sqrt{2}\kappa e^{a_{p+4}\bphi/2}}
  \Bigg\{ \bar{\varphi} - \varphi\big(\BRP,\TQ_{p+4}\big)
  \Bigg[ 1 - \left( 1 - \frac{b\big(\BRP,\TQ_p,\TQ_{p+4}\big)}{\bar{b}} \right) \times \nonumber\\
  && \left( \frac{2}{3-p}\frac{1}{\BDM} +
  \frac{\BDM-\BDP}{\BDS\BDP+\BDS\BDM-2\BDP\BDM} \right) \Bigg] \Bigg\} \nonumber\\
  &=& - \frac{\bbeta V_p}{\sqrt{2}\kappa e^{a_p\bphi/2}}
  \left[ \bar{\varphi} - \varphi\big(\BRP,\TQ_{p+4}\big) \right] ,
  \label{eq:dIGC-dQpf}
\end{eqnarray}
where
\begin{eqnarray}
  \bar{\varphi} &=& \sqrt{2} \kappa e^{a_{p+4} \bphi/2} \bar\Phi_{p+4} ,\nonumber\\
  \varphi\big(\BRP,\TQ_{p+4}\big) &=& \left( \frac{\BDP}{\BDM} \right)^{1/2} \left( \frac{1-\BDM}{1-\BDP} \right)^{1/2} .
  \label{eq:def-varphi}
\end{eqnarray}
Again in getting this result we have used (\ref{eq:reduced-charge-density})
and the equation of state. Setting this derivative to zero would give
exactly the second expression in (\ref{eq:phi-explicit}). In the same
fashion, if we partially differentiate $\IGG$ with respect to $\BRP$, $\Q_p$
and $\Q_{p+4}$, we would obtain all the three equations in
(\ref{eq:local-temperature}) and (\ref{eq:phi-explicit}).

For future simplifications in the computation, we define some reduced
quantities in the following.
The reduced action in the CC ensemble is defined as
\begin{eqnarray}
  \RICC &\equiv&  \frac{\kappa^2 \ICC}{2\pi \brho_b^{4-p} V_{p+4} \Omega_{4-p}} \nonumber\\
  &=& - \bar{b} \left[ 2 \left( \frac{\BDP}{\BDM} \right)^{1/2}
  + (3-p) (\BDP\BDM)^{1/2} \left( 1+ \frac{1}{\BDS} \right) + 2p - 8 \right] \nonumber\\
  && - (1-\BDP)^{1/2} \left(1- \frac{\BDP}{\BDM} \right)^{\frac{1}{3-p}}
  \left( 1- \frac{\BDP}{\BDS} \right)^{1/2} .
  \label{eq:reduced-ICC}
\end{eqnarray}
Other reduced quantities can be defined as follows,
\begin{eqnarray}
  x \equiv  \left( \frac{\brho_+}{\brho_b} \right)^{3-p} < 1 , \qquad
  Q \equiv  \TQ_p < 1 ,\qquad
  q \equiv  \left( \frac{\TQ_{p+4}}{\brho_b} \right)^{3-p} < x .
\end{eqnarray}
Accordingly,  in the reduced variables, $\bar\Delta_{+,-,*}$ and the other
reduced actions can be expressed as
\begin{eqnarray}
  \BDP &=&  1 - x ,\nonumber\\
  \BDM &=&  1 - \frac{q^2}{x} ,\nonumber\\
  \BDS &=&  \frac{\BDP + \BDM + \sqrt{\displaystyle (\BDM-\BDP)^2 + 4 Q^2 \BDP\BDM}}{2(1-Q^2)} ,
  \label{eq:reduced-delta}
\end{eqnarray}
and
\begin{eqnarray}
  \RIGC &=& \RICC - (3-p) \bar{b} Q \bar{\Phi} ,\nonumber\\
  \RICG &=& \RICC - (3-p) \bar{b} q \bar{\varphi} ,\nonumber\\
  \RIGG &=& \RICC - (3-p) \bar{b} Q \bar{\Phi} - (3-p) \bar{b} q \bar{\varphi} .
  \label{eq:other-reduce-action}
\end{eqnarray}
Notice that $\bar \Delta_+$ only depends on $x$, and $\bar \Delta_-$
depends on both $q$ and $x$, while $\bar \Delta_*$ depends on all three
variables $x$, $q$ and $Q$. 
 With (\ref{eq:reduced-ICC}) and (\ref{eq:other-reduce-action}), we find again the equations of equilibrium state
corresponding to (\ref{eq:dICC-dBRP}), (\ref{eq:dIGC-dQp}) and
(\ref{eq:dIGC-dQpf}), in reduced quantities,
\begin{eqnarray}
  \PP{\RIGC}{Q} &=& -(3-p) \bar{b} \left\{ \bar{\Phi} - \Phi(x,Q,q)
  \left[ 1 + \left( 1 - \frac{b(x,Q,q)}{\bar{b}} \right)
  \frac{\BDS(\BDM-\BDP)}{\BDS\BDP+\BDS\BDM-2\BDP\BDM} \right] \right\} ,\nonumber\\
  \PP{\RICG}{q} &=& -(3-p) \bar{b} \left\{ \bar{\varphi} - \varphi(x,q)
  \left[ 1 - \left( 1 - \frac{b(x,Q,q)}{\bar{b}} \right) \left( \frac{2}{3-p}\frac{1}{\BDM}
  + \frac{\BDM-\BDP}{\BDS\BDP+\BDS\BDM-2\BDP\BDM} \right) \right] \right\} ,\nonumber\\
  \PP{\RICC}{x} &=& f(x,Q,q) \big[ \bar{b} - b(x,Q,q) \big] ,
  \label{eq:stationary-point-equations}
\end{eqnarray}
where the functions $f$, $b$, $\Phi$ and $\varphi$ have been defined in (\ref{eq:def-b-f}),
(\ref{eq:def-Phi}) and (\ref{eq:def-varphi}). Notice that from
(\ref{eq:deltas}), we find $0<\Phi<1$ and $0<\varphi<1$ for nonzero
$p$- and $(p+4)$-brane charges when the horizon is not coincident with the
boundary.

\section{Thermodynamics in different ensembles\label{sect:Ensembles}}

Note that when $Q=0$, the whole system becomes a \Dppf-brane system,
and 
the various phase structures of this system have already been thoroughly
analyzed in \cite{lu:2011,lu:2011-2}. Thus, in the
following calculations, we will always assume that $Q>0$ ($\Phib>0$).
Nevertheless, we will still compare our results with the $Q=0$ case
for consistency check or revealing the different traits in the
presence of \Dp-branes.

\subsection{Overview of the behaviors of $b(x)$}
\label{sec:typical-curves}

Before performing concrete analyses in  specific ensembles, we first
examine a few typical behaviors of the function $b(x,Q,q)$, which are
useful in our later exploration.

In the following discussions of the phase structures in these ensembles,
the key problems to be solved are finding 
out the stationary points of thermodynamic potentials and which one is 
stable. That is, we need to solve the following equations,
\begin{eqnarray}
  b(x,Q,q) &=& \bar{b} ,\nonumber\\
  \Phi(x,Q,q) &=& \Phib ,\nonumber\\
  \varphi(x,q) &=& \vphib ,
  \label{eq:stationary-equation}
\end{eqnarray}
and then determine whether the solution(s) to these equations is
the minimum point of those  thermodynamic potentials. In CC ensemble,
we only need the first equation. In CG ensemble, we need the first and
the third ones and in GC ensemble, we need the first and the second ones.
In GG ensemble, we need all the three equations. For simplicity, in
this paper we constraint ourselves only to the stability of branes under
the change of the horizon size as in \cite{Chamblin:1999tk,lu:2011-2,Wu:2011yu}.
As will be shown later, this is equivalent to the thermal stability condition
that the specific heat capacity is positive for the equilibrium state.
We will also give a brief discussion in a later section on the other
electrical stability conditions like those used in \cite{Chamblin:1999hg}
and leave the details in a paper to appear \cite{xiao:2015}.
So here we will assume that the second and third equations in (\ref{eq:stationary-equation})
have already been satisfied\cite{lu:2011-2,Wu:2011yu}, and $Q$ and $q$
can be solved and be substituted in the first equation.
Thus, the only problem left is to solve the first equation where $b$
is a function of just one variable $x$, that is,
\begin{eqnarray}
  b(x) \eval{\Phi=\Phib\,\,\text{or/and}\,\, \varphi=\vphib} = \bar{b},
  \label{eq:simplified-stationary-equation}
\end{eqnarray}
for corresponding ensembles.  In that case, the local minimum
condition of the thermodynamical potential would simply be
\begin{eqnarray}
  \DD{^2I}{x^2} > 0 .
  \label{eq:minimum-condition}
\end{eqnarray}

In equation (\ref{eq:simplified-stationary-equation}), $\bar{b}$ is
the inverse temperature on the boundary which is assumed to be a
constant input parameter. Hence finding out the solution is equivalent
to finding out the intersection points of curve $b=b(x)$ and the horizontal
line $b=\bar{b}$. On the other hand, we will demonstrate that
\begin{eqnarray}
  \DD{^2I}{x^2} \sim - \DD{b}{x} ,
  \label{eq:D2I-Dx2-Db-Dx}
\end{eqnarray}
which means the local minimum point is the intersection point where
$\DD{b}{x}$ (the slope) is negative.

\begin{figure}[!ht]
  \centering
  \includegraphics[width=0.45\textwidth]{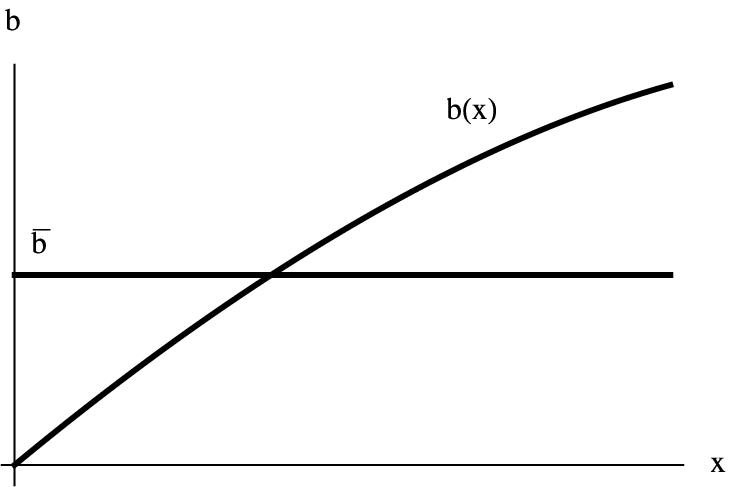} \quad
  \includegraphics[width=0.45\textwidth]{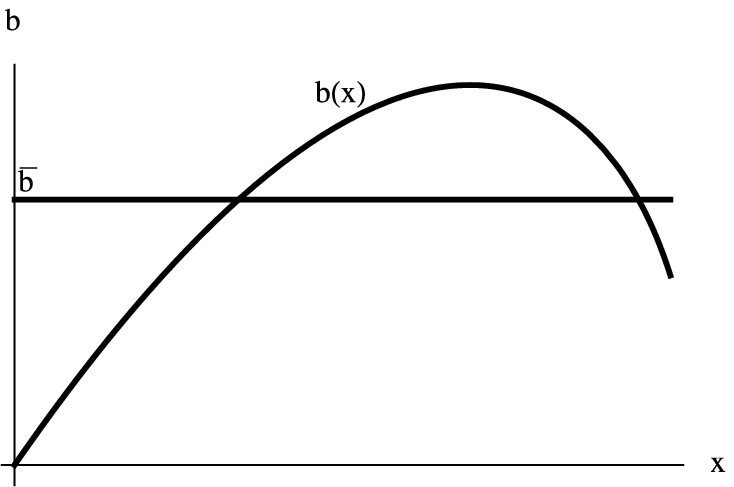}
  \caption{Typical behaviors of $b(x)$ in GG ensembles as well as in some cases in $GC$ and $CG$ ensemble.}
  \label{fig:GG-typical}
\end{figure}

We will see in the GG ensemble, that
there are only two kinds of curves  for $b(x)$ which are listed in
Figure~\ref{fig:GG-typical}.
One can see that, in the first graph of Figure~\ref{fig:GG-typical},
although there is an intersection point of the two curves, it is not
a minimum of grand potential since the slope at that point is positive.
We therefore conclude that there is no stable black brane phase for
any $\bar{b}$ given that $b(x)$ increases monotonically. In this case,
since the hot flat space can have the same boundary condition with the
brane, the only stable phase should be the hot flat space.  However, in
the second graph, for $\bar{b}$ within some range, there could be two
intersection points, where the slope is negative at the one with  larger $x$. Thus we
conclude that there will be a locally stable black brane phase 
for $\bar{b}$ within that range, i.e., the brane configuration with
larger horizon is locally stable. To find out whether it
is  a globally stable phase, we have to compare
the grand potential at this point with the one for the hot flat space,
i.e. the one at $x=0$. If the grand potential at this point is a global minimum it would be the real
stable phase, otherwise  it is merely a locally stable one. In this
section, we only concentrate on  locally stable phases.

\begin{figure}[!ht]
  \centering
  \includegraphics[width=.45\textwidth]{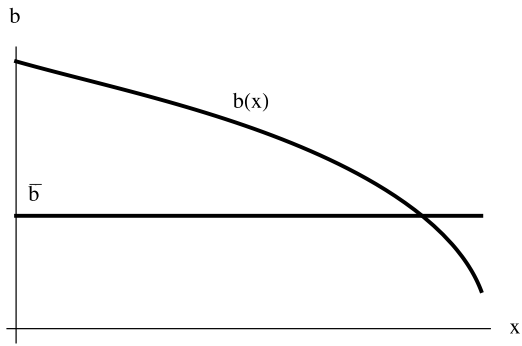}
  \caption{A typical curve in GC ensemble for $p=1$ besides those in
Figure~\ref{fig:GG-typical}.}
  \label{fig:GC-p-1}
\end{figure}

\begin{figure}[!ht]
  \centering
  \includegraphics[width=.45\textwidth]{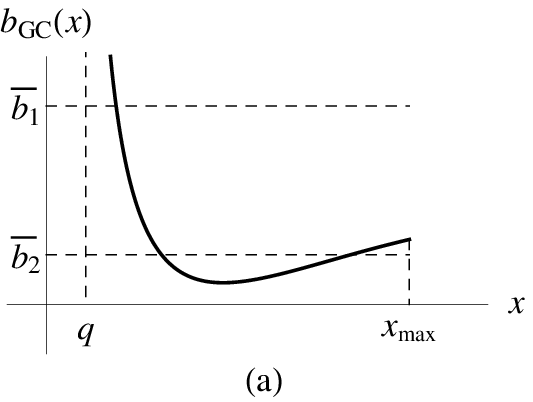} \quad
  \includegraphics[width=.45\textwidth]{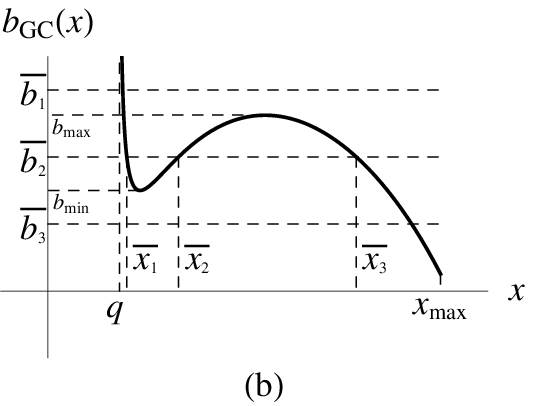}
  \includegraphics[width=.45\textwidth]{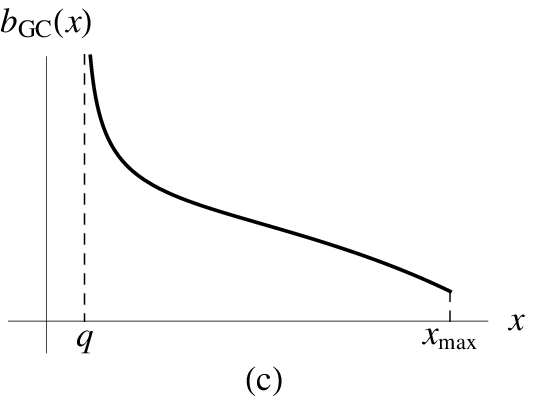}
  \caption{Typical $b(x)$ behaviors in GC ensemble for $p=0$}
  \label{fig:GC-p-0}
\end{figure}

In the GC ensemble, there will be
much richer phase structures. For $p=2$ (D2-D6 system), the possible curves of $b(x)$
are the same as the ones for the GG ensemble, and therefore we do not need to
reanalyse them. For $p=1$ case (D1-D5 system), there will be one more possible
shape besides the two appearing in Figure~\ref{fig:GG-typical}. This
new curve monotonically decreases as shown in Figure~\ref{fig:GC-p-1}.
Thus, if the constant $\bar{b}$ line intersects with $b(x)$ curve at some point,
this point must correspond to a globally stable black brane phase. If $p=0$
(D0-D4 system),
there are three cases as shown in Figure~\ref{fig:GC-p-0}. The 
third one as shown in Figure~\ref{fig:GC-p-0}(c) is similar to  Figure~\ref{fig:GC-p-1} with decreasing $b(x)$. 
In Figure~\ref{fig:GC-p-0}(a), if $\bar{b}=\bar{b}_1$ ($\bar b>b_{GC}(x_{max})$),
there is only one intersection point at which the state is stable,
whereas if $\bar{b}=\bar{b}_2$ ($\bar b<b_{GC}(x_{max})$), there will be two intersection points
and only the one with smaller $x$ is locally stable. For the second 
graph,
there are more possibilities. We can see that if
$\bar{b}=\bar{b}_1>b_{max}$
or $\bar{b}=\bar{b}_3<b_{min}$, where $b_{min}$ and $b_{max}$ are the
local minimum and local maximum of $b(x)$ as shown in the graph, there is only one intersection point and it
is a stable black brane phase. Yet for $\bar{b}=\bar{b}_2$ with $b_{min}<\bar b_2 <b_{max}$, there can be three
intersection points. Denoting these three points as $x_1$, $x_2$
and $x_3$ with $x_1<x_2<x_3$, one can see that the point with $x=x_2$ is apparently
unstable for its positive
slope while the other two are both locally stable. As regards these two
locally stable phases,  the one with higher free
energy will eventually transit to the other phase. It can be shown 
that there exists some value of $\bar{b}$
between  $b_{min}$ and  $b_{max}$, at which the two
locally stable phases have equal thermodynamic potentials but
different entropies, which indicates a first order phase transition,
i.e. the van der Waals-like phase transition. The second graph in
Figure~\ref{fig:GC-p-0} is just a typical case of this kind which is
similar to the one-charge black brane. In two-charge case
$b(x_{max})$ does not reach zero, so there can be possibilities that
$b(x_{max})>b_{min}$. We will discuss this case in section
\ref{sec:GC-phase}.

In the CG ensemble, all three possible $b(x)$ curves in Figure \ref{fig:GC-p-0} also appear here. As argued in \cite{lu:2012-2}, in the CC
ensemble, the delocalized \Dp-branes and the \Dppf-branes are
equipotent in the sense that exchanging the two kinds of  brane
charges would yield similar phase structures. This reveals some
symmetry under exchange of the boundary conditions that we have imposed on these
two kinds of branes. The GC and CG ensemble are just related by an
exchange of the boundary conditions on these two kinds of branes and
we will see later that the phase structures of these two ensembles are
really related in this way.
 
Now  having found all the patterns of the $b(x)$ curves and  the
locally stable phase in each case qualitatively, we need to be specific in
each ensemble to find out in what ranges of the parameter $Q$ (or
$\Phib$) and $q$ (or $\vphib$) one can have a certain pattern of
$b(x)$. Then, we would know the possible state of the system  given
any pair of $Q$ (or $\Phib$) and $q$ (or $\vphib$). The following
subsections are devoted to this problem.

The CC ensemble (especially when $p=1$) has already been explored
by Lu et al. in \cite{lu:2012-2}, and we have gathered some results in
Appendix~\ref{ap:CC} including some facts not given in
\cite{lu:2012-2}.  Interested  readers are refered to their paper
for more details. In the following subsections we
will focus on the other three kinds of ensembles. 

\subsection{GG ensemble}

As stated in the previous section, we adopt the one variable analysis
as an embryo attempt to find out the thermodynamical structure.
We need to solve $Q$ and $q$ in terms of $x$ and the corresponding electric potentials
on the boundary ($\bar \Phi$ and $\bar \varphi$), from the
two electromagnetic equilibrium equations, i.e. the second and the
third equations in (\ref{eq:stationary-equation}),  using (\ref{eq:def-Phi}) and
(\ref{eq:def-varphi}).
Since the $Q$ and $q$ dependence  comes from $\BDM$ and $\BDS$, it is convenient to
first express 
$\BDM$ and $\BDS$ in terms of $x$, $\Phib$ and $\vphib$,
\begin{eqnarray}
  \BDM &=& \frac{\BDP}{\xi} ,\qquad
  \BDS = \BDP \frac{1-\Phib^2}{\displaystyle \xi-\Phib^2} ,
  \label{eq:GG-delta-x}
\end{eqnarray}
where $\xi=1-(1-\vphib^2)x<1$. Then the charges can easily be
obtained,
\begin{eqnarray}
q=\frac {x\bar\varphi}{\xi^{1/2}}\,,\qquad
Q=\frac{\bar\Phi(1-\xi)}{(1-\bar\Phi^2)\xi^{1/2}}\,.
\label{eq:q-Q}
\end{eqnarray}
The reduced action which is proportional to the grand potential can then be obtained in terms of $x$, $\bar\varphi$
and $\bar \Phi$,
\begin{eqnarray}
  \RIGG
  = 2(4-p)(1-\sqrt\xi)[\bar b-b_0(x)]
  \label{eq:IGG}
\end{eqnarray}
where
\begin{eqnarray}
  b_0(x) \equiv \frac{1+\xi^{\frac12}}{2(4-p)}
  \left( \frac x{1-\bar\Phi^2} \right)^{\frac12}
  (1-\xi)^{\frac {1-p} {2(3-p)}} .
  \label{eq:CG-b0-x}
\end{eqnarray}
We have mentioned in (\ref{eq:deltas}) that
$  \RP^{3-p} > \RM^{3-p} > k$.
 This
inequality can be rewritten in terms of $\BDP$, $\BDM$ and $\BDS$,
\begin{eqnarray}
  \BDP < \BDM \ , \quad \BDM < \BDS .
  \label{eq:new-ordering}
\end{eqnarray}
The first relation in (\ref{eq:new-ordering}) is guaranteed by the first
equality of (\ref{eq:GG-delta-x}).  The second relation holds only for
\begin{eqnarray}
  x < \frac{1-\Phib^2}{1-\vphib^2} .
  \label{eq:GG-x-upper-bound}
\end{eqnarray}
So, we have a restriction on $x$,
\begin{eqnarray}
  0 < x < x_{max}
  = \textrm{min}\left\{1,\frac{1-\Phib^2}{1-\vphib^2}\right\} .
  \label{eq:GG-x-restriction}
\end{eqnarray}
For $\bar\Phi<\bar \varphi$, $x_{max}=1$ which means the horizon should
be inside the boundary. For $\bar\Phi>\bar \varphi$, the condition
is the requirement of $Q<1$. This limit should not be reached since
otherwise $\Delta_{*}$ would blow up and the size of the time
direction and some space dimensions
would shrink to zero. In fact, before it shrinks to string scale,
the quantum effects should be large, and the supergravity
approximation is invalid. Thus, if at $x_{max} $ the system has lower
grand potential than at the local minimum, we will regard the system
as unstable, since either the horizon tends to the boundary or the
supergravity approximation is not  applicable.  At $x=0$, the charges
tends to zero, the system reduce to  the ``hot flat space''.

Next, we need to find out the stationary point by setting
\begin{eqnarray}
  0 &=& \PP{\RIGG(x,\bar\Phi,\bar \varphi)}{x}\bigg|_{\bar\Phi,\bar\varphi}
  = f_{GG}(x) \left[ \bar{b} - b_{GG}(x) \right] ,
  \label{eq:GG-stationary-condition}
\end{eqnarray}
where
\begin{eqnarray}
  b_{GG}(x) &=&  \frac{1}{3-p} \left(\frac{x\xi}{1-\Phib^2}\right)^{1/2}
  \Big(1-\xi\Big)^{-\frac{1-p}{2(3-p)}} ,\nonumber\\
  f_{GG}(x) &=& (4-p) \frac{1-\xi}{x \xi^{1/2}} > 0 .
  \label{eq:GG-func-b-f}
\end{eqnarray}
Equation (\ref{eq:GG-stationary-condition}) again reduces to the
familiar one,
\begin{eqnarray}
  \bar{b} = b_{GG}(x) .
  \label{eq:GG-reduced-stationary-condition}
\end{eqnarray}
Notice that at $x=0$, $b_{GG}$ is always zero. So
$b_{GG}$ should be increasing 
near $x=0$ due to the fact that $b_{GG}\geq 0$, i.e. $d
b_{GG}(x)/dx\ge0$ at $x=0$. Finally
the local stability condition is effectively
\begin{eqnarray}
  \DD{^2\RIGG(\bar{x})}{x^2} \sim - \DD{b_{GG}(\bar{x})}{x} > 0 ,
  \label{eq:GG-stability-condition}
\end{eqnarray}
where $\xb$ is the solution to equation
(\ref{eq:GG-reduced-stationary-condition}). The derivative of
$b_{GG}$ can be evaluated,
\begin{eqnarray}
  \DD{b_{GG}}{x} = \frac{b_{GG}}{2(3-p)x\xi} \left[ 2- (5-p)(1-\vphib^2)x \right] .
  \label{eq:GG-db-dx}
\end{eqnarray}
Thus, the condition (\ref{eq:GG-stability-condition}) is equivalent to
\begin{eqnarray}
  2- (5-p)(1-\bar{\varphi}^2) \bar{x} < 0 ,
  \label{eq:GG-simplified-condition}
\end{eqnarray}
which is exactly the same condition as for black $(p+4)$-brane without
$p$-brane charge, i.e. the (35) in \cite{lu:2011-2},
except that now the solution $\xb$ depends on a new parameter $\bar
\Phi$.

From (\ref{eq:GG-db-dx}), we see that there is a turning point
\[ x_0 = \frac{2}{5-p} \frac{1}{1-\vphib^2} \]
where $b_{GG}(x)$ increases for $x<x_0$ and decreases for $x>x_0$.
Therefore, if $x_{max}<x_0$, $b_{GG}(x)$ will be a monotonically
increasing function for $x\in(0,x_{max})$, otherwise it will have
a maximum at $x=x_0$. For the former case (referred to as \emph{case A}), the curve of $b_{GG}$
looks like the one in the first graph of Figure~\ref{fig:GG-typical},
while for the latter case (referred to as \emph{case B}) it looks like the curve in the second
graph, just as we stated in the previous section.

\begin{figure}[!ht]
  \centering
  \includegraphics[width=.5\textwidth]{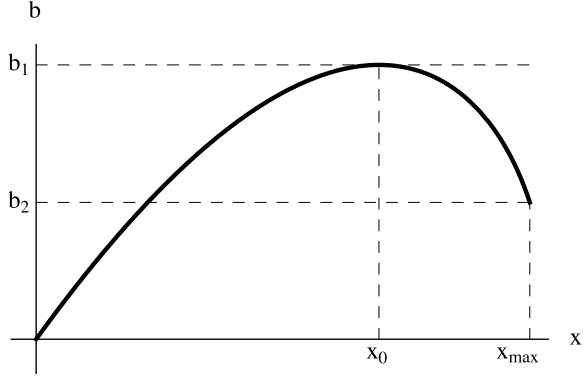}
  \caption{$b_{GG}$ for $x_0<x_{max}$. For $\bar\Phi<\bar \phi$,
  $x_{max}=1$, otherwise $x_{max}=\frac{1-\Phib^2}{1-\vphib^2}$.}
  \label{fig:GG-shape-n}
\end{figure}

We then need to figure out the condition for $\Phib$ and $\vphib$ in 
each case. We start by requiring $x_0>x_{max}$, i.e.,
\begin{eqnarray}
  \frac{2}{(5-p)(1-\vphib^2)} >
  \textrm{min}\left\{1,\frac{1-\Phib^2}{1-\vphib^2}\right\} .
  \label{eq:GG-x0-xmax}
\end{eqnarray}
which would give
\begin{eqnarray}
  \Phib > \sqrt{\frac{3-p}{5-p}} \quad \textrm{or} \quad
  \vphib > \sqrt{\frac{3-p}{5-p}} .
  \label{eq:GG-Phi-phi-mono}
\end{eqnarray}
That means that for $\Phib$ and $\vphib$ satisfying (\ref{eq:GG-Phi-phi-mono}),
case A is applied and the system cannot have a  
stable black brane phase. 
 On the contrary, if
\begin{eqnarray}
  \Phib < \sqrt{\frac{3-p}{5-p}} \quad \textrm{and} \quad
  \vphib < \sqrt{\frac{3-p}{5-p}} ,
  \label{eq:GG-Phi-phi-n}
\end{eqnarray}
$b_{GG}(x)$ will look like the curve in Figure~\ref{fig:GG-shape-n}.
One can see that if $b_1>\bar{b}>b_2$, there can be a locally stable black brane phase and its
horizon radius is between $x_0$ and $x_{max}$, where $b_1$ and $b_2$
are defined by
\begin{eqnarray}
  b_1 &=& b_{GG}(x_0) = \frac{ 2^{\frac 1{3-p}} (5-p)^{-\frac{5-p}{2(3-p)}} }
  { \sqrt{(3-p)(1-\bar\Phi^2)(1-\bar\varphi^2)}} ,\nonumber\\
  b_2 &=& b_{GG}(x_{max}) = \left\{ \begin{matrix}
    \frac{\vphib (1-\vphib^2)^{\frac{1-p}{2(3-p)}}}{(3-p)\sqrt{1-\Phib^2}}
    \,, \text{ for } \Phib < \vphib ;\cr
    \frac{\Phib(1-\Phib^2)^{\frac{1-p}{2(3-p)}}}{(3-p)\sqrt{1-\vphib^2}}
    \,, \text{ for } \vphib < \Phib \,.
  \end{matrix} \right.
  \label{eq:b1-b2}
\end{eqnarray}
For $\bar{b}$ not in this range, there will be no stable black brane phase.

\begin{figure}[!ht]
  \centering
  \includegraphics[width=.6\textwidth]{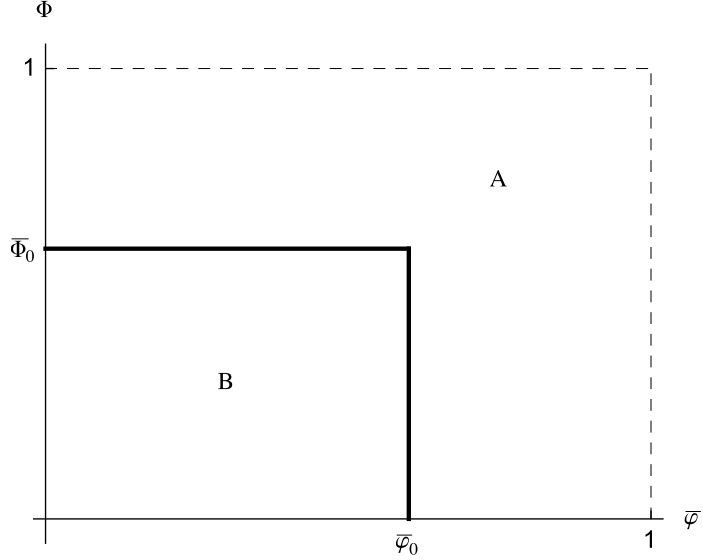}
  \caption{Parameter plane in GG ensemble}
  \label{fig:GG-diagram}
\end{figure}

Combining the arguments above, we would have the diagram 
shown in Figure~\ref{fig:GG-diagram} which gives a more explicit view
of the answer to the previous question.
In this figure, the two constants are
\begin{eqnarray}
  \Phib_0 = \vphib_0 = \sqrt{\frac{3-p}{5-p}} .
  \label{eq:GG-phi-0}
\end{eqnarray}

\begin{figure}[!ht]
  \centering
  \includegraphics[width=0.4 \textwidth]{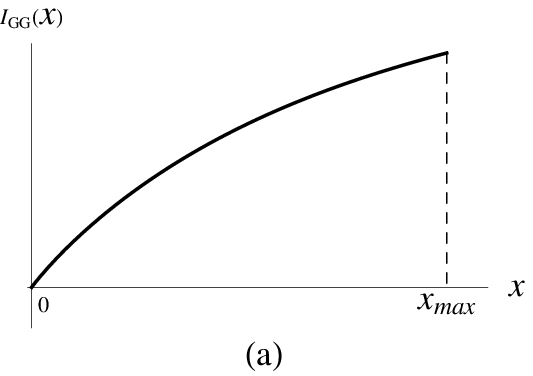}\quad
  \includegraphics[width=0.4 \textwidth]{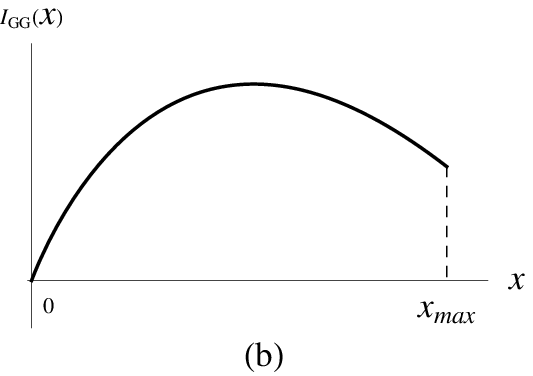}\quad
  \includegraphics[width=0.4 \textwidth]{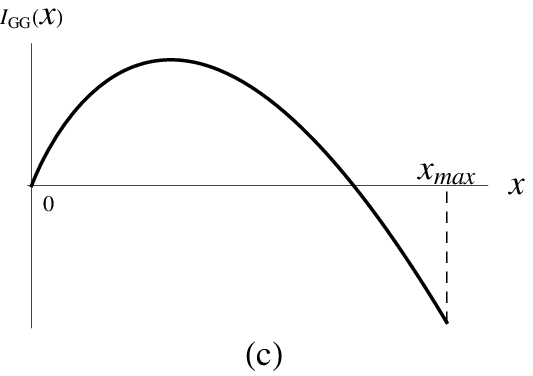} \\
  \includegraphics[width=0.4 \textwidth]{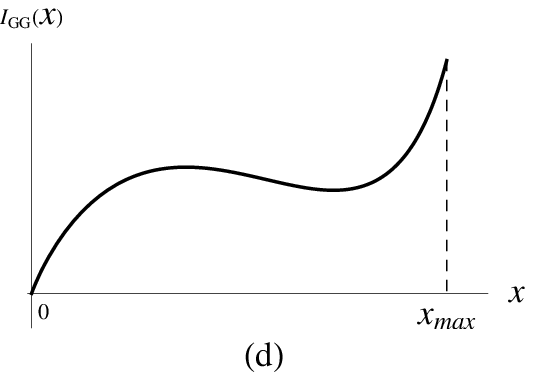}\quad
  \includegraphics[width=0.4 \textwidth]{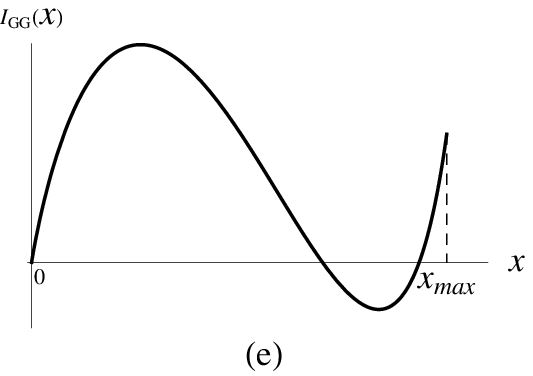}\quad
  \caption{Typical $I_{GG}(x)$ behaviors in region A and B. }
  \label{fig:IGG}
\end{figure}

Since the charges are not conserved in grand canonical ensemble, the
hot flat space can be a possible phase. In region A, at very low
temperatures
with $\bar b>b_2$ where $\partial I_{GG}/\partial{x}>0$,
the grand potential is a monotonically increasing function of
$x$ as shown in Figure~\ref{fig:IGG}(a) (we will use the subfigure
labels to denote different cases). Since the hot
flat space has $x=0$, the final stable phase should be the hot flat
space. If $\bar b<b_2$, there can be two cases which are
shown in (b) and (c)  in Figure~\ref{fig:IGG}. In the first
case (b), which has lower temperature than (c),
the minimum of the action is still at $x=0$ which corresponds to the hot
flat space. For case (c) which has higher temperature, the global minimum of
the $I_{GG}$ is at the boundary $x_{max}$ which means that the hot flat space is
not a global minimum of the grand potential and hence is unstable, and the
horizon of the black brane tends to expand to the boundary. This means
that there is no stable phase in this region. We can find out the
condition for this case by solving the inequality $I_{GG}(x_{max})<0$ which gives 
\begin{eqnarray}
  \bar b<b_{\rm unstable}=
  \left\{ \begin{array}{l}
    \frac{(1-\Phib^2)^{\frac{1}{3-p}}}{2(4-p)} \sqrt{\frac{1+\Phib}{(1-\Phib)(1-\bar\varphi^2)}}
    \,,\text{ for } \bar\Phi>\bar\varphi \\
    \frac{(1-\vphib^2)^{\frac{1}{3-p}}}{2(4-p)} \sqrt{\frac{1+\vphib}{(1-\vphib)(1-\Phib^2)}}
  \,,\text{ for } \bar\varphi>\bar\Phi
  \end{array} \right.
  \label{eq:b-unstable}
\end{eqnarray}
So for temperatures larger than $1/b_{\rm unstable}$ the system is unstable. 
We do not know what really happens at this high temperature.  Below
this temperature, only hot flat space is the stable phase. Region B,
where $\max(\bar\Phi,\bar\varphi)<\sqrt{\frac {3-p}{5-p}}$, has more
cases to be considered. First, for very low temperatures, similar to
case A, when there is no solution for
(\ref{eq:GG-stationary-condition}), $I_{GG}$ behaves as Figure~\ref{fig:IGG}(a)
and the stable phase is the hot flat space. For higher temperatures when
(\ref{eq:GG-stationary-condition}) has two solutions, i.e. $b_2<\bar
b<b_1$, $I_{GG}$ has two stationary points, one unstable and the other
locally stable. The possible behaviors of $I_{GG}$ are shown in (d) and (e) in
Figure~\ref{fig:IGG}. For case (d) which has lower temperature, the
hot flat space at $x=0$ has the lowest grand potential, and hence is
the stable phase. At higher temperatures such that the system
corresponds to graph (e), the grand potential at the locally stable
point is negative and becomes the global minimum. So in this case the
final stable phase is the black brane. For certain $\bar \Phi$ and
$\bar \varphi$, case (e) can not happen. We can
find out the condition for (e) to happen by looking at
$I_{GG}(\bar x)=0$. We put the detailed analysis in Appendix \ref{ap:GG} and
only state the results here. Firstly, when 
\begin{eqnarray}
  \max(\bar \Phi,\bar \varphi)>\frac{3-p}{5-p}\,,
  \label{eq:GG-Phi-phi-1}
\end{eqnarray}
for $\bar b>b_2$, $I_{GG}(\bar x)$
is always positive which corresponds to graph (d). Under this
circumstance, the cases corresponding to graph (e) does not happen.
With such $\bar \Phi$ and $\bar \varphi$, when $\bar
b\in (b_{\rm unstable},b_2)$, there is only one solution to
(\ref{eq:GG-stationary-condition}), which
also corresponds to (b) in Figure~\ref{fig:IGG}. Hot flat space is
then
the global minimum as in case A. For higher temperatures such that
$\bar b<b_{\rm unstable}$, graph (c)
also appears and the system is unstable.  Secondly, when 
\begin{eqnarray}
  \max(\bar \Phi,\bar \varphi)<\frac{3-p}{5-p}\,,
  \label{eq:GG-Phi-phi-2}
\end{eqnarray}
there is a temperature $T_0\in(1/b_1,1/b_2)$ at which the locally stable
point is at $\bar x_0(\bar
\varphi)\equiv \frac {4(4-p)}{(5-p)^2(1-\bar\varphi^2)}$. If the
temperature is lower than $T_0$ such that at the locally stable point
$\bar x<\bar x_0(\bar \varphi)$, $I_{GG}$ is positive (case (d)) and
the hot flat space is the globally stable phase. Only for temperature
higher than $T_0$, when the locally stable point $\bar x > \bar
x_0(\bar\varphi)$, the black brane has a negative grand potential,
which corresponds to  graph (e) in Figure~\ref{fig:IGG}, and black brane is the final globally stable phase. For much higher
temperature such that $\bar b < b_2$, the case
corresponding to graph (c) also happens, and the system is again unstable.
So combined with case A, for all
$\max(\bar\Phi,\bar\varphi)>\frac{3-p}{5-p}$, the black brane can not
be the final stable phase and for low temperatures the hot flat space
is the global minimum. Only when
$\max(\bar\Phi,\bar\varphi)<\frac{3-p}{5-p}$, the black brane can be
the final stable phase for certain temperature  $T\in(T_0,1/b_2)$.

As we stated in the previous section, from the conditions
(\ref{eq:GG-Phi-phi-mono}, \ref{eq:GG-Phi-phi-n}, \ref{eq:b1-b2},
\ref{eq:GG-phi-0}), we see that the phase structure is symmetric under
the exchange of the boundary conditions imposed on the two kinds of
branes, i.e. exchanging $\bar \Phi$ and $\bar\varphi$. We will also
find similar results in GC and CG ensemble.

\subsection{GC ensemble}
\label{sec:GC-phase}

In this ensemble, the \Dppf-brane charge $q$ is fixed, therefore we
only need to use the equation
\begin{eqnarray}
  \Phi(x,Q) = \Phib
  \label{eq:GC-charge-equilibrium}
\end{eqnarray}
to get rid of the unfixed quantity $Q$. Since $\Phib$ is fixed at the
boundary, we can solve $\bar\Delta_*$  from  this equation and then
$Q$ using (\ref{eq:delta-star}),
\begin{eqnarray}
  \BDS &=& \frac{\BDP\BDM(1-\bar{\Phi}^2)}{\displaystyle \BDP-\BDM\bar{\Phi}^2} ,
  \label{eq:GC-delta-star-x}
\end{eqnarray}

\begin{equation}
  Q_{GC} = \frac{\bar{\Phi} (\BDM-\BDP)}{(1-\bar{\Phi}^2) (\BDP\BDM)^{1/2}} .
  \label{eq:GC-Q-x}
\end{equation}
Using (\ref{eq:GC-delta-star-x}) and also requiring $\BDP<\BDM<\BDS$, we
can find the domain of $x$,
\begin{eqnarray}
  0 < q < x < x_{max} < 1 ,
  \label{eq:GC-x-range}
\end{eqnarray}
where
\begin{eqnarray}
  x_{max} = \frac{\displaystyle 1-\bar{\Phi}^2 +
\sqrt{(1-\bar{\Phi}^2)^2+4q^2\bar{\Phi}^2}}{2}\, .
  \label{eq:GC-x-max}
\end{eqnarray}
As in GG ensemble, when $x\to x_{max}$, $\Delta_{*}$ tends to infinity and
could not be reached in supergravity approximation.
At $x=q$, at which $\Delta_+=\Delta_-=\Delta_*$ and $Q_{GC}=0$, the system reduces
to extremal $(p+4)$-brane. However, the horizon of the extremal brane
is singular and the quantum gravity effect must be non-negligible,
and hence the Euclidean action method is not
applicable. So if the global minimum of the reduced action is at the
boundary $x=q$ or $x_{max}$, we do not know what the final stable state
of the system is. So, we regard the system as unstable in  our semi-classical
description when the thermodynamical
potential at $x=q$ or $ x=x_{max}$ is lower than the local minimum
value.

With relation (\ref{eq:GC-Q-x}), we can find the stationary point by
solving equation
\begin{eqnarray}
  0 &=& \DD{\RIGC}{x} = f_{GC}(x) \left[ \bar{b} - b_{GC}(x) \right] ,
  \label{eq:GC-first-condition}
\end{eqnarray}
where
\begin{eqnarray}
  b_{GC}(x) &=& b(x,Q_{GC}(x,q,\bar\Phi),q)=\frac
1{3-p}\left(1-\frac{\Delta_+}{\Delta_-}\right)^{\frac{p-1}{2(3-p)}}\sqrt{\frac
{\Delta_+(1-\Delta_+)}{\Delta_-(1-\bar\Phi^2)}} ,\nonumber\\
  f_{GC}(x) &=& \frac{(3-p) \BDM (\BDM-\BDP) + (5-p)
  (\BDP+\BDM-2\BDP\BDM)}{2 \BDP^{1/2} \BDM^{3/2} (1-\BDP)} > 0\,,
  \label{eq:GC-b-f}
\end{eqnarray}
which gives the equilibrium condition $\bar b=b_{GC}(x)$ as  before.
Assuming that $x=\xb$ solves (\ref{eq:GC-first-condition}),
the stability condition becomes
\begin{eqnarray}
  \DD{b_{GC}(\xb)}{x} < 0 .
  \label{eq:GC-reduced-condition}
\end{eqnarray}
Now we compute the derivative of $b_{GC}$,
\begin{eqnarray}
  \DD{b_{GC}}{x}\bigg|_{q,\bar\Phi} = \frac{b_{GC}}{2x} \left( 1 +
  \frac{(\BDP+\BDM-2\BDP\BDM)[2\BDP-(3-p)\BDM]}{(3-p)\BDP\BDM(\BDM-\BDP)}
  \right) .
  \label{eq:GC-db-dx}
\end{eqnarray}
We can see that in this expression, in the large parentheses, $\BDS$
completely disappears, which means the equation is independent of $\Phib$.
For $\Phib=0$, i.e. D$(p+4)$-brane without D$p$-brane  charges, which is equivalent to setting $\BDS=\BDM$ in
$b_{GC}(x)$, (\ref{eq:GC-db-dx}) would exactly recover  the
same  equation for D$(p+4)$-brane in canonical ensemble
((46) in \cite{lu:2011}). This is also true even for $\bar\Phi\neq0$
since it is independent of $\bar \Phi$,
that is,  (\ref{eq:GC-db-dx}) still gives  the same condition
to determine the signature of
$\displaystyle\frac{db_{CG}}{dx}$ for $\bar \Phi=0$, which was
already given as  in (47) of \cite{lu:2011}. This means that the stationary
point for $b_{GC}(x)$ is independent of $\bar \Phi$ and only depends
on the charge $q$. To be specific, we
rewrite (\ref{eq:GC-db-dx}) in terms
of $x$ and $q$,
\begin{eqnarray}
  \DD{b_{GC}}{x} &=& - g(x,q)
  \frac{b_{GC}}{x^4(1-x) (1-\frac{q^2}{x}) (1-\frac{q^2}{x^2})} ,
  \label{eq:GC-db-dx-x-q}
\end{eqnarray}
where
\begin{eqnarray}
  g(x,q) &=& \frac{5-p}{2} x^4 - \left( 1+ \frac{7-p}{2}q^2 \right) x^3
  - \frac{3-3p}{2} q^2 x^2 \nonumber\\
  && + q^2 \left( 2-p+ \frac{9-3p}{2}q^2 \right) x - (3-p)q^4 .
  \label{eq:func-g}
\end{eqnarray}
Now  (\ref{eq:func-g}) is exactly the function on
the left hand side of (47) in \cite{lu:2011} (there, a different
parameter, the ``co-dimension'' $\tilde{d}$ which is equal to $3-p$ in
our case, is used). Therefore, condition (\ref{eq:GC-reduced-condition})
becomes a simpler one,
\begin{eqnarray}
  g(\bar{x},q) > 0 .
  \label{eq:GC-simplified-condition}
\end{eqnarray}
With (\ref{eq:GC-first-condition}) and (\ref{eq:GC-simplified-condition})
we can proceed with the phase structure analysis for different $p$.

\begin{figure}[!ht]
  \centering
  \includegraphics[width=.6\textwidth]{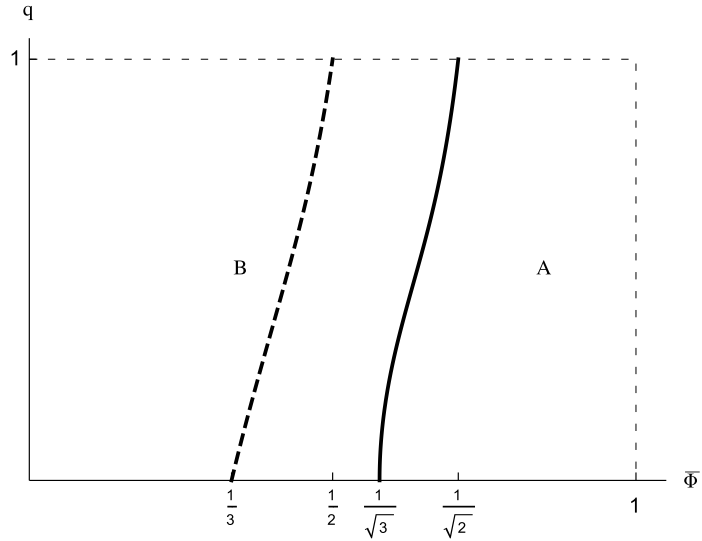}
  \caption{$q-\bar \Phi$ parameter plane in D2-D6 GC ensemble. Region B includes
all the region on the left of the solid curve. }
  \label{fig:GC-diagram-p-2}
\end{figure}

First for $p=2$ (D2-D6 system), notice that at $x=q$, $b_{GC}=0$ which is
the same as $b_{GG}$ in GG ensemble at $x=0$. As we have mentioned in
section~\ref{sec:typical-curves}, there are only two possible
shapes of $b_{GC}(x)$ which are similar to the cases in the GG ensemble
(see Figure~\ref{fig:GG-typical}) with the only difference that the
left end point of the $b_{GC}(x)$ is at $x=q$ here.  Now we will find out
the conditions for these two cases. From Figure~\ref{fig:GG-typical} one
can see that the difference of the two patterns is the sign of the
slope of the curves at $x_{max}$. So one can conclude that there must
be a transition line on the $\Phib-q$ plane on which the $b_{GC}(x)$
curve has $db_{GC}(x)/dx =0 $  at  $x_{max}$.  Actually this is a universal
feature for all $p$, so instead of solving it here for only $p=2$ case
we have done it for all $p$ and put the tedious analytic computations
in Appendix \ref{ap:GC}.  Here we just give the diagram in
Figure~\ref{fig:GC-diagram-p-2} for $p=2$, which has similar meaning
as Figure~\ref{fig:GG-diagram}.
In region A, $b_{GC}$ is a monotonically increasing function and there
is no locally stable black brane phase. In region B, the $b_{GC}(x)$ curve is like 
Figure~\ref{fig:GG-shape-n} and for $b_2<\bar{b}<b_1$, there are two
intersection points and the larger one is 
a locally stable black brane phase. The boundary between A and B on $q-\bar \Phi$ plane has the following form,
\begin{eqnarray}
  q = \frac{1-\Phib^2}{\displaystyle \Phib(3-5\Phib^2)}
  \sqrt{2(1-\Phib^2)(3\Phib^2-1)} ,
  \quad \frac{1}{\sqrt{3}} < \Phib < \frac{1}{\sqrt{2}} .
  \label{eq:GC-boundary-p-2}
\end{eqnarray}
The locally stable state may not be a global minimum of the reduced
action because at the boundary $x=q$  the reduced action may
have smaller value. To find out the condition that the local minimum can not be
the global minimum for $b_2<\bar b<b_1$ one can look at the critical
case by requiring
that at $\bar b=b_{GC}(x_{max})$ the $I_{GC}(x_{max})=I_{GC}(x=q)$. This
equation can be solved analytically to give 
\begin{eqnarray}
q=\frac {2(1-\bar\Phi)^2(3\bar\Phi-1)}{\Phib(3-5\bar\Phi)}\,,
  \quad \frac{1}{3} < \Phib < \frac{1}{2} .
\label{eq:q-Phi-p-2}
\end{eqnarray} 
which is denoted as the dashed curve in Figure~\ref{fig:GC-diagram-p-2}.
On the right of this curve, when $b_2<\bb<b_1$, we always have
$I_{GG}(x_{max})>I_{GG}(\xb)>I_{GG}(q)$, hence the minimum of reduced action is at $x=q$.
When $\bb<b_2$ the global minimum is either at $x=q$ or
$x=x_{max}$. There is a critical temperature $T_0=1/b_0>1/b_2$ where the
reduced action at $x=q$ and $x=x_{max}$ are equal. Below this
temperature the global minimum is at $x=q$ while above it is at $x=x_{max}$.
In both cases, we regard the system as unstable. On the left of the dashed curve, whether the
locally stable state becomes the global minimum of the reduced action
depends on the temperature. There is a temperature denoted as
$T_{\rm unstable}\equiv1/b_{\rm unstable}$ with $b_2<b_{\rm unstable}<b_1$ such
that the reduced action at the locally stable point equals the one
at $x=q$, i.e. $I_{GC}(\bar x)=I_{GC}(q)$. Below this temperature,
i.e.
$\bar b>b_{\rm unstable}$, the global minimum is still at $x=q$. Only for
$b_2<\bar b<b_{\rm unstable}$, the local stability becomes a global
one. For higher temperatures with $\bar b < b_2$, the global
minimum of the reduced action is at $x=x_{max}$ and the system is
unstable again.

\begin{figure}[!ht]
  \centering
  \includegraphics[width=0.4 \textwidth]{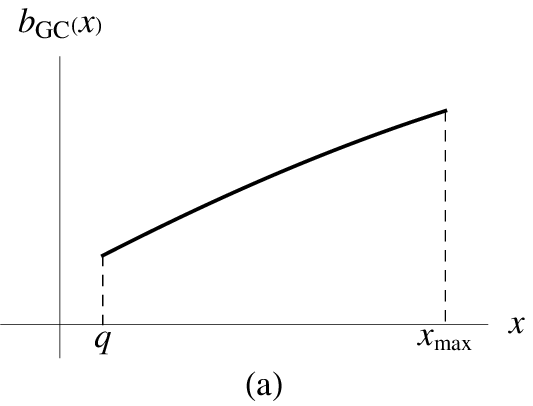}\qquad
  \includegraphics[width=0.4 \textwidth]{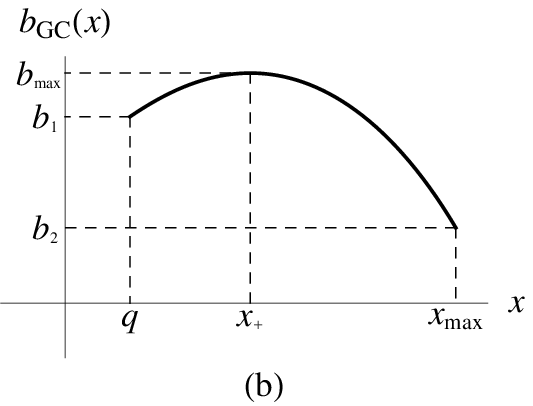}\qquad
  \includegraphics[width=0.4 \textwidth]{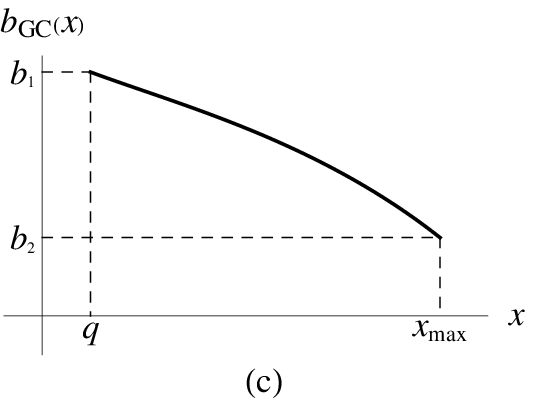}
  \caption{Typical $b_{GC}(x)$ behaviors for D1-D5 system. (a)
$q<1/3$ and $x_{max}<x_+$, corresponds to region A; (b) $q<1/3$ and
$x_{max}>x_+$, corresponds to region B; (c)$q>1/3$, corresponds to
Region C.}
  \label{fig:bGC-p-1}
\end{figure}
\begin{figure}[!ht]
  \centering
  \includegraphics[width=.6\textwidth]{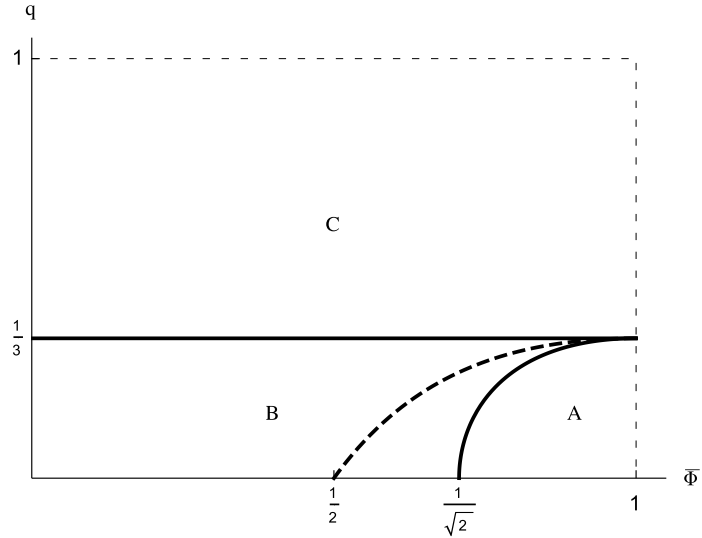}
  \caption{Parameter plane in D1-D5 GC ensemble. Region B includes
all the region on the left of the solid line between A and B. The
$b_{GC}$ behaviors of A, B, C correspond to (a), (b), (c) in Figure
\ref{fig:bGC-p-1}\,.}
  \label{fig:GC-diagram-p-1}
\end{figure}

Next for $p=1$,  at the left end $x=q$, $b_{GC}(x=q)$ has a non-vanishing
finite limit which is
different from GG ensemble and $p=2$ case. Now, $g(x,q)$ can be factorized as
\begin{eqnarray}
  g(x,q) = (x+q)(x-q)(x-x_-)(x-x_+) ,
  \label{eq:GC-g-x-p-1}
\end{eqnarray}
where
\begin{eqnarray}
  x_\pm = \frac{1+3q^2 \pm \sqrt{(1-q^2)(1-9q^2)}}{4} .
  \label{eq:GC-x-pm-p-1}
\end{eqnarray}
One can see that for $q>1/3$, which is region C in Figure~\ref{fig:GC-diagram-p-1}, $x_-$ and $x_+$ are complex conjugates
and $g(x,q)>0$, which means $b_{GC}$ decreases
monotonically for $x\in(q,x_{max})$ as shown in Figure~\ref{fig:bGC-p-1}(c). In this case, there is a globally stable black brane phase
for $b_2<\bar b<b_1$. For $q<1/3$, it is easy to verify
$x_-<q<x_+$, and therefore $g(x,q)$ is negative ($b_{GC}$ increases) for
$x\in(q,x_+)$ and is positive ($b_{GC}$ decreases) for $x>x_+$. This
indicates that if $x_+>x_{max}$, $b_{GC}$ is monotonically increasing
(Figure~\ref{fig:bGC-p-1}(a)) and there is no stable black brane
phase. This case is denoted as region A in Figure~\ref{fig:GC-diagram-p-1}. For 
$x_+<x_{max}$, the behavior of $b_{GC}$ is shown in
\ref{fig:bGC-p-1}(b) and there exists a locally stable black brane phase for $b_2<\bar{b}<b_{max}$,
which is denoted as region B in Figure~\ref{fig:GC-diagram-p-1}. 
 On the boundary
between A and B, $q$ and $\Phib$ are related by
\begin{eqnarray}
  q = \frac{\sqrt{\displaystyle (2-\Phib^2)(2\Phib^2-1)}}{3\Phib} ,
  \quad \frac{1}{\sqrt{2}} < \Phib < 1 .
  \label{eq:GC-boundary-p-1}
\end{eqnarray}
In region B, like in $p=2$ case, the locally stable black brane state
may not be the global minimum and the $x=q$ point may have lower
reduced action. The condition that the system is just to have only
$x=q$ as the global minimum of the reduced action for all $\bar b >b_2$ is
\begin{eqnarray}
q=\frac 1 3(5-2\bar \Phi-\frac 2 {\bar\Phi}), \quad \frac 1
2<\bar\Phi<1\,,
\label{eq:q-Phi-p-1}
\end{eqnarray}
which is also denoted as the dashed curve in Figure~\ref{fig:GC-diagram-p-1}. 
On the right of the curve, the global minimum is either at
$x=q$ for $\bar b >b_2$  or at $x=x_{max}$ for $\bar b <b_2$ and the
system is regarded as unstable.
On the left of the dashed curve, there is a possibility that at some
temperature $T_{\rm unstable}\equiv 1/b_{\rm unstable}$ the reduced action at $x=q$ equals the one at the
locally stable point (see Appendix \ref{ap:dashed-lines} for further details on how to
find this temperature). Below this temperature ($\bar b>b_{\rm unstable}$),
the reduced action at $x=q$ is the global minimum and the system tends
to $x=q$ similar to the cases in region C and the system is also
unstable.  The locally stable one
becomes a globally stable one only when the temperature is above
$T_{\rm unstable}$, i.e. $b_2<\bar b <b_{\rm unstable}$. When the temperature
is higher than $1/b_2$, $x=x_{max}$ becomes the global minimum of the
reduced action, and so, the system becomes unstable again. 

\begin{figure}[!ht]
  \centering
  \includegraphics[width=.6\textwidth]{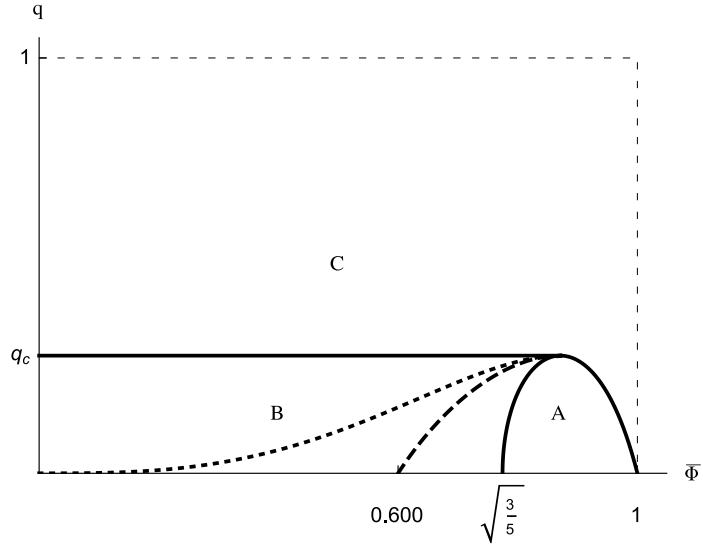}
  \caption{Parameter plane in D0-D4 GC ensemble. Region B includes
all the region on the left of the solid line between A and B. The
$b_{GC}$ behaviors of A, B, C correspond to (a), (b), (c) in Figure
\ref{fig:GC-p-0}\,.}
  \label{fig:GC-diagram-p-0}
\end{figure}

Finally, for the $p=0$ case, there are  three kinds of $b_{GC}(x)$
behaviors which are  shown in Figure~\ref{fig:GC-p-0}. Similar to the
one-charge black $(p+4)$-brane case \cite{lu:2011}, there exists a
critical charge $q_c$ beyond which $b_{GC}(x)$ is a monotonically
decreasing function for arbitrary $\bar\Phi$ as shown in Figure
\ref{fig:GC-p-0}(c). When $q=q_c$ and $0<\bar \Phi<\bar\Phi_{max}$,
where $\bar \Phi_{max}$ is obtained in Appendix \ref{ap:GC} to be $
\Phib_{max} \cong 0.871417$, $b_{GC}$ has an inflection point at which
a second order phase transition occurs.  At $\bar\Phi_{max}$, the
position of the 
critical point $\bar x$ is equal to $x_{max}$. For $\bar \Phi>\bar
\Phi_{max}$ at $q=q_c$, the critical point position $\bar x$ is
larger than $ x_{max}$ and can not be
reached by the supergravity solution and hence $b_{GC}$ is also
a monotonically decreasing function. So, in $q-\Phi$ plane,
the region denoted as C  in
Figure~\ref{fig:GC-diagram-p-0}, which contains only decreasing
$b_{GC}(x)$, includes the $q>q_c$ region as well as a small region
with $q<q_c$ where the solutions for $d
b_{GC}(x)/dx=0$ are larger than $x_{max}$. 
In this figure, $q_c\approx 0.141626$  is the same critical value as
the $\tilde{d}=3$ case in \cite{lu:2011} as expected.  So, in region
C, there is always a stable black brane phase for $\bar b>
b_{GC}(x_{max})$.  In region A and B, curves of $b_{GC}$ have the same
shape as in the other two diagrams (a) and (b) in
Figure~\ref{fig:GC-p-0}, respectively. On the boundary lines between A
and B and between A and C, $q$ and $\Phib$ has following relation,
\begin{eqnarray}
  q = \frac{1-\Phib^2}{\displaystyle \Phib(9-7\Phib^2)}
  \sqrt{2(3-\Phib^2)(5\Phib^2-3)} ,
  \quad \sqrt{\frac{3}{5}} < \Phib < 1 .
  \label{eq:GC-boundary-p-0}
\end{eqnarray}
 In every region (A, B, C) there are chances for this system to have a
locally stable black brane phase.  In all these three cases, as $x\to
q$,  $db_{GC}/dx<0$ which means that the locally stable black brane
can exist for arbitrarily small temperatures. But the minimum of
$b_{GC}$ does not tend to zero which means that at high enough
temperature the locally stable black brane can not exist. In  region C,
at  temperatures higher than $1/b_{GC}(x_{max})$, the global minimum is
at $x_{max}$. In region A, there is a temperature $T_{\rm unstable}$ higher than
$1/b_{GC}(x_{max})$, at which the reduced action at $x_{max}$ is equal to
the one at the local minimum. When the temperature is
higher than $T_{\rm unstable}$ the global minimum is at $x_{max}$.
Under both
these two circumstances the system is unstable and we do not know
what the final state is. So, in region A and region C, at temperature
below $1/b_{GC}(x_{max})$ and $T_{\rm unstable}$, respectively, the black brane
is the globally stable phase. In region B  which corresponds to Figure
\ref{fig:GC-p-0}(b), there can also be a first order phase transition
under some circumstances. As stated in section
\ref{sec:typical-curves}, we need to discuss two cases separately,
$b_{min}>b_{GC}(x_{max})$ and $b_{min}<b_{GC}(x_{max})$ which
correspond to the left and right regions of the dotted line in region
B, respectively.  When
$b_{min}>b_{GC}(x_{max})$, the system has one locally stable black brane phase for
$b_{GC}(x_{max})<\bar b< b_{min}$ or $\bar b >b_{max}$, which is also
the global minimum.  For  $ b_{min}<\bar{b}< b_{max}$, there can be
two locally stable black brane phases and the final phase should be the one with lower
reduced action. There is a first-order phase transition temperature
$T_t=1/b_t$ at which the reduced actions of the two phases are equal.
Above (or below) this temperature the larger (or the smaller) one is
the global minimum. This is the same as the corresponding case in
canonical ensemble of one-charge black brane. When
$b_{min}<b_{GC}(x_{max})$, we still define a temperature $T_t\equiv 1/ b_t$, at which
the reduced action at the two local minima are equal even when one
minimum has $\bar x>x_{max}$ . In this case, we need to consider
two subcases.  The first
one has $b_t>b_{GC}(x_{max})$.  Then when $\bar b< b_{GC}(x_{max})$, i.e. the
temperature higher than $1/ b_{GC}(x_{max})$, the minimum of the
reduced action is at $x=x_{max}$ and the system is unstable.  Below this temperature, the analysis is the
same as the above case and the system can have a first order transition at
$T_t$ between the two locally stable minima.  The second subcase is
when $b_t< b_{GC}(x_{max})$. In this subcase, at $b_t$ the larger
solution for $b_t=b_{GC}(x)$ goes beyond $x_{max}$ and there is
another temperature $T_{\rm unstable}\equiv 1/b_{\rm unstable}(>T_t)$ at which $I_{GC}(x_{max})$ equals the one at the
local minimum with smaller $\bar x$.  Then, condition for 
$x_{max}$ to be the global minimum of the reduced action is $\bar
b<b_{\rm unstable}$. So
for temperature higher than $T_{\rm unstable}$ the system is unstable and we can not say much about the phase of
the system. For $b_{GC}(x_{max})>\bar b>b_{\rm unstable}$, the smaller locally
stable black brane  has the lower reduced action than at $x=x_{max}$,
and hence is the global minimum. So for temperature lower than
$T_{\rm unstable}$, the smaller locally stable black brane is the global minimum.
The condition for $b_t=b_{GC}(x_{max})$ can only be solved numerically and
is  denoted as the dashed line in Figure~\ref{fig:GC-diagram-p-0}.
On the dashed line, $b_{\rm unstable}$ coincides with $b_t$ and
$b_{GC}(x_{max})$.
The left region of the dashed line corresponds to the $b_t>
b_{GC}(x_{max})$ case which has the van der Waals-like phase structure
and the right of the line in B corresponds to the $b_t<
b_{GC}(x_{max})$ case.
 
\subsection{CG ensemble}

In this ensemble, we fix the \Dp-brane charge $Q$ and the potential $\vphib$
for D$(p+4)$ at the boundary. Using the
``electromagnetic'' equilibrium equation,
\begin{eqnarray}
  \varphi(x,q) = \bar{\varphi} ,
  \label{eq:CG-charge-equilibrium}
\end{eqnarray}
 the reduced action of the system depends
on the only variable $x$,
\begin{eqnarray}
  \RICG(x) = \RICG(x,Q={\rm const.\,},q=\qCG(x,\bar\varphi)) ,
  \label{eq:CG-2-variable-x-q}
\end{eqnarray}
where $\qCG$ is the solution to (\ref{eq:CG-charge-equilibrium}),
\begin{eqnarray}
  \qCG(x,\bar\varphi) = \frac{x \bar{\varphi}}{\xi^{1/2}} .
  \label{eq:func-q-x}
\end{eqnarray}
This relation also guarantees that $x>\qCG$ is satisfied as long as $\vphib<1$. 
At $x\to 0$, $q_{CG}\to 0$ and $\xi\to 1$. 
From (\ref{eq:GG-delta-x}), we can see that $\bar \Delta_- \to
\BDP$, 
 which means the curvature
singularity is coming closer to the horizon. Actually it can be shown that
the scalar curvature blows up at the horizon $x=0$ by explicit calculations.
The quantum effect near horizon, therefore, may be  large such that  the
semi-classical approach can not be applied. So, as in the GC ensemble, if at $x=0$ or
$x=1$, the system has lower reduced
action than the local minimum, we will consider it as unstable in the
semi-classical approach.

Now we can perform the one-variable analysis by requiring
\begin{eqnarray}
  \DD{\RICG(x)}{x} = f_{CG}(x) \left[ \bar{b} - b_{CG}(x) \right] = 0 ,
  \label{eq:CG-first-condition}
\end{eqnarray}
where
\begin{eqnarray}
  b_{CG}(x) &=& b(x,Q,\qCG) ,\nonumber\\
  f_{CG}(x) &=& \frac{\BDM-\BDP}{2 \BDP^{1/2} \BDM^{1/2} (1-\BDP)}
  \left[ 5-p + (3-p) \frac{\BDM(\BDM-\BDP)}{\BDS\BDP + \BDS\BDM - 2\BDP\BDM} \right] > 0 .\quad
  \label{eq:CG-def-b-f}
\end{eqnarray}
Let us assume $\xb$ to be the solution of equation
(\ref{eq:CG-first-condition}), then the minimum condition of free energy
is again reduced to
\begin{eqnarray}
  \DD{b_{CG}(\xb)}{x} < 0 .
  \label{eq:CG-reduced-condition}
\end{eqnarray}
The left hand side of the above inequality is
\begin{eqnarray}
  \DD{b_{CG}(x)}{x} = \frac{b_{CG}}{2(3-p)x} \left[ 2- \frac{(3-p)\BDM}{\BDP}
  + \frac{(3-p)\BDM(\BDM-\BDP)}{\BDS\BDP+\BDS\BDM-2\BDP\BDM} \right] .
  \label{eq:CG-db-dx}
\end{eqnarray}
One can easily check that when $Q=0$, i.e. $\BDS=\BDM$,
(\ref{eq:CG-db-dx}) automatically falls back to equation (34) in
\cite{lu:2011-2}. Now with (\ref{eq:CG-db-dx}) and the first equation
in (\ref{eq:CG-def-b-f}), we can redo all the analyses done in the
two previous subsections. However, since the computations are complicated
and tedious, we put the details into Appendix \ref{ap:CG}
for interested readers. Here we just list the final results.

\begin{figure}[!ht]
  \centering
  \includegraphics[width=.6\textwidth]{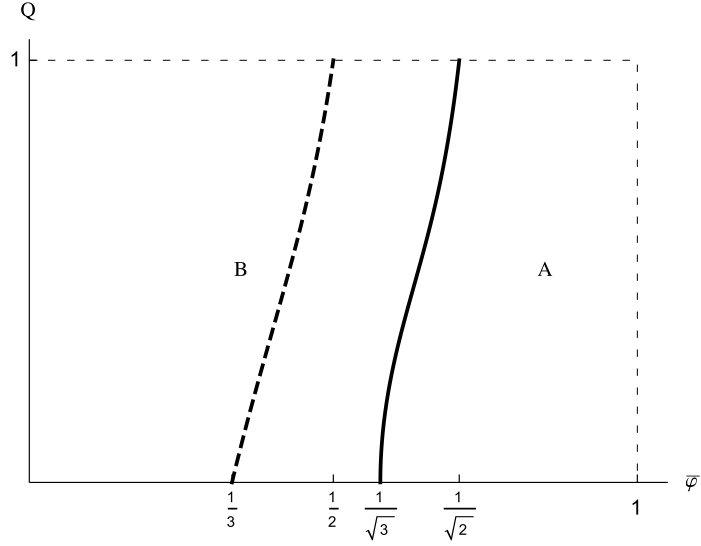}
  \caption{$Q-\bar\varphi$ parameter plane in D2-D6 CG ensemble}
  \label{fig:CG-diagram-p-2}
\end{figure}

For $p=2$ case, similar to  Figure~\ref{fig:GC-diagram-p-2} in GC 
ensemble, we have the following graph (Figure~\ref{fig:CG-diagram-p-2}).
In area A, since $b_{CG}$ increases monotonically
as the left graph in  Figure~\ref{fig:GG-typical}, there is no stable black brane phase,
while in area B, $b_{CG}$ behaves as in  Figure
\ref{fig:GG-shape-n},
there  exists a locally stable black brane state for $b_2<\bar b
<b_1$. The curve separating region A and B is described by
\begin{eqnarray}
Q=\frac{(-1+\bar\varphi^2)\sqrt{2(1-\bar\varphi^2)(3\bar\varphi^2-1)}}{\bar\varphi(5\bar\varphi^2-3)},
\quad \frac{1}{\sqrt 3}<\bar \varphi<\frac{1}{\sqrt 2}.
\label{eq:CG-boundary-p-2}
\end{eqnarray}
Again, the local minimum may or may not be the
global minimum. It should compete with the boundary point $x=0$ or
$x=1$. Similar to the discussion in GC ensemble, in
Figure~\ref{fig:CG-diagram-p-2}, on the right of the dashed line, the
local minimum can not be the global minimum of the system and on the
left, it can be the global minimum for certain temperature
$T\in (T_{\rm unstable},1/b_2)$, where $T_{\rm unstable}$ is the temperature at
which the reduced action at the local minimum equals the one at $x=0$.
All the discussions in the GC ensemble can be used here and we would
not repeat them.  The dashed line is described by
\begin{eqnarray}
  Q = \frac
  {2(1-\bar\varphi^2)(3\bar\varphi^2-1)}{ \bar\varphi(3-5\bar\varphi^2)} ,
  \quad \frac{1}{3} < \bar\varphi < \frac{1}{2} .
  \label{eq:Q-phi-p-2}
\end{eqnarray}
Notice if one makes the exchanges $\bar \Phi\leftrightarrow \bar \varphi$ and
$q\leftrightarrow Q$, (\ref{eq:CG-boundary-p-2}) and
(\ref{eq:Q-phi-p-2})  exchange with
(\ref{eq:GC-boundary-p-2}) and (\ref{eq:q-Phi-p-2}).
So, Figure~\ref{fig:CG-diagram-p-2} is the same as
Figure~\ref{fig:GC-diagram-p-2}  except the labels.

\begin{figure}[!ht]
  \centering
  \includegraphics[width=.6\textwidth]{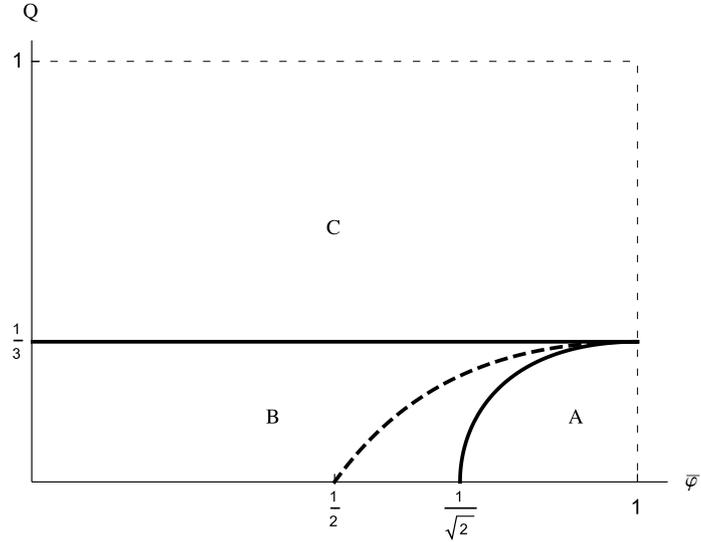}
  \caption{\sf Parameter plane in D1-D5 CG ensemble}
  \label{fig:CG-diagram-p-1}
\end{figure}

For $p=1$ case, the behaviors of $b_{CG}(x)$ are the same as the ones in
GC ensemble and Figure~\ref{fig:bGC-p-1} can also be used here with the
only difference that the domain is now $0<x<1$. In the $Q-\bar
\varphi$ plane, the three cases of (a), (b), (c) in Figure~\ref{fig:bGC-p-1}
correspond to region A, B, C in Figure~\ref{fig:CG-diagram-p-1}.
In the lower right region A, $b_{CG}$ is monotonically increasing and
therefore there is no stable black brane phase. In the upper half
region C, since $b_{CG}(x)$ is always decreasing, there is always a
locally stable black brane state for
$b_{CG}(0)>\bar{b}>b_{CG}(1)$ when constant $\bar b$ has an
intersection point with $b_{CG}$ curve. In the lower left region B,
for $b_{CG}(1)<b<b_{max}$, where $b_{max}$ is the maximum of $b_{CG}$,
 there will be a locally stable black brane
phase since the constant $\bar{b}$ line intersects with $b_{CG}$ at
some point where $db_{CG}/dx<0$. The curve separating region A and region B is solved in the
appendix to be 
\begin{eqnarray}
  Q = \frac{\sqrt{\displaystyle
(2-\bar\varphi^2)(2\bar\varphi^2-1)}}{3\bar\varphi} ,
  \quad \frac{1}{\sqrt{2}} < \bar\varphi < 1 .
  \label{eq:CG-boundary-p-1}
\end{eqnarray}

For the local minimum in B to be a global minimum, the system should be in
the region on the left of the dashed line. On the right of the dashed
line, the global minimum is either at $x=0$ or at $x=1$ and hence the
system is unstable. On the left of the dashed
line,  there is a $ b_{\rm unstable}$ between $b_{CG}(1)$ and the
maximum $b_{max}$ of
$b_{CG}$, where the local minimum of the reduced action equals the
reduced action at $x=0$.  Only for temperatures such that $b_{CG}(1)<\bar b
< b_{\rm unstable}$, the local minimum becomes
the global minimum. The dashed line is described
by 
\begin{eqnarray}
Q=\frac 1 3(5-2\bar \varphi-\frac 2 {\bar \varphi})\,,
\quad \frac 1
2<\bar\varphi<1\,,
\label{eq:Q-phi-p-1}
\end{eqnarray}
One may have already found that (\ref{eq:CG-boundary-p-1}) and
(\ref{eq:Q-phi-p-1}) are the same as (\ref{eq:GC-boundary-p-1}) and
(\ref{eq:q-Phi-p-1}) except the exchanges of $q\leftrightarrow Q$ and
$\bar \Phi\leftrightarrow \bar \varphi$. 
All the discussions of different phases
are the same as the ones in GC ensemble for $p=1$ and we will not
repeat them here.

\begin{figure}[!ht]
  \centering
  \includegraphics[width=.6\textwidth]{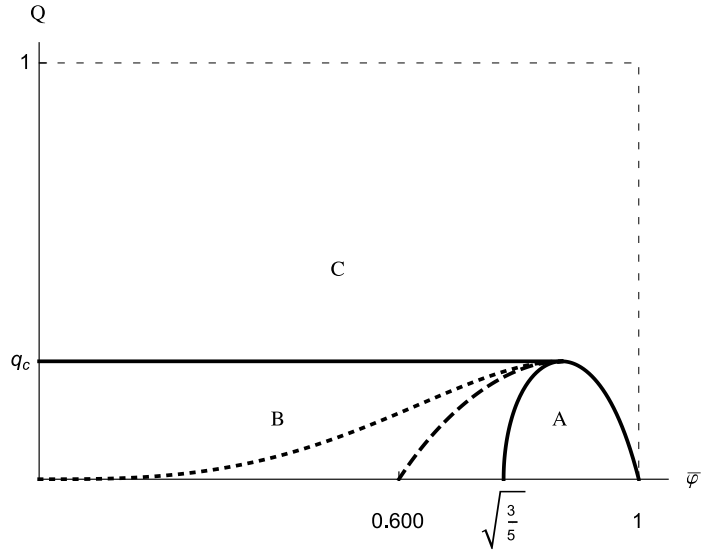}
  \caption{Parameter plane in D0-D4 CG ensemble}
  \label{fig:CG-diagram-p-0}
\end{figure}

The $Q-\bar \varphi$  plane graph for $p=0$ case is shown in Figure~\ref{fig:CG-diagram-p-0}.
In all three regions, since as $x\to 0$, $b_{CG}$ tends to
positive infinity, there is always a stable black brane phase at 
arbitrarily low temperature. As long as $\bar\varphi$ is not zero,
$b_{CG}$ has a finite minimum, which means for high enough
temperatures, the black brane can not exist. To be specific, in region
C which occupies almost all valid area of the parameter plane, the
behavior of $b_{CG}$ is decreasing  similar to Figure
\ref{fig:GC-p-0}(c) with the domain of $x$ replaced by $0<x<1$.  So
there is always a locally stable black brane phase for $\bar{b}>b_{CG}(1)$.  In region
A, which takes only a very small portion of the plane, $b_{CG}$
behaves like Figure~\ref{fig:GC-p-0}(a) which decreases first and then
increases. Thus there could be a locally stable black brane phase if that solution $\xb$
is in the decreasing segment.  In region B, the $b_{CG}$ curve looks
like Figure~\ref{fig:GC-p-0}(b), and as discussed in the GC ensemble,
there can be a van der Waals-like phase transition on the left of the
dashed line in Figure~\ref{fig:CG-diagram-p-0} when the phase
transition temperature $T_t=1/b_t$ is lower than $1/b_{CG}(1)$,
where $b_t$ can be obtained by requiring the reduced actions to be equal at
the two local minimum. On the right side of the dashed line, similar
to what we did in the GC ensemble, we define a temperature $T_{\rm unstable}(>T_t)$
at which the reduced action at the local minimum with smaller $\bar x$
equals  the one at $x=1$. In this case, below the temperature
$T_{\rm unstable}$, the smaller black brane is the global minimum. For
temperature higher than $T_{\rm unstable}$ the global minimum tends to
$x=1$ which means the instability of the system and  a failure of
our method. A second order phase transition can happen on the boundary
line between regions B and C,  when $Q=Q_c\cong 0.141626$ and
$\vphib<0.871417$.  On this line, $b_{CG}$ has an inflection point where a
second order phase transition  occurs and the two phases
 are indistinguishable. All the detailed discussions are
the same as the $p=0$ case in GC ensemble.

As we have mentioned before, the three diagrams in this subsection
are essentially the same as those in the previous subsection except
that we substitute $\Phib$ and $q$ for  $\vphib$
and $Q$. Even the critical charges are precisely equal, $q_c=Q_c$.
This indeed supports the statement that the \Dp\ branes and the \Dppf\
branes are equipotent as far as only the thermodynamics is concerned,
which has been pointed out in \cite{lu:2012-2}. One may think of it
as a symmetry,
\begin{eqnarray}
  \vphib \leftrightarrow \Phib ,\quad Q \leftrightarrow q.
  \label{eq:symmetry}
\end{eqnarray}

\section{More general thermodynamic stability
conditions\label{sect:General-stability}}
The stability condition (\ref{eq:minimum-condition}) for the thermodynamic potential we used here
is the one with respect to the fluctuation of the horizon size. It can
be shown that this condition is the same as the positivity of the
specific heat capacity for each ensemble as follows:
According to the first thermodynamic law and our definitions of the
potentials, for equilibrium state
$dE=TdS+(3-p)\Phi dQ+(3-p)\varphi dq$ , one can find
$\left(\frac{\partial E}{\partial x}\right)_{Qq}
=T(x,Q,q)\left(\frac{\partial S}{\partial x}\right)_{Qq}$
with $T(x,Q,q)=1/b(x,Q,q)$, which can also be checked by explicit calculation.
For CC ensemble, from $\tilde I_{CC}=\bar{b}E(x,Q,q)-S(x,Q,q)$ and
by using the equilibrium condition 
\begin{eqnarray}
  0=\left(\frac{\partial I_{CC}}{\partial x}\right)_{Qq}=\bar{b}
  \left( \frac{\partial E}{\partial x} \right)_{Qq}
  - \left( \frac{\partial S}{\partial x} \right)_{Qq}
  = (\bar{b}-b) \frac{1}{b} \left( \frac{\partial S}{\partial x} \right)_{Qq}
\end{eqnarray}
Comparing with  (\ref{eq:stationary-point-equations}), one finds that
$\frac 1 b\left(\frac{\partial S}{\partial x}\right)_{Qq}
=\frac 1 b\left(\frac{\partial S}{\partial b}
\frac{\partial b}{\partial x}\right)_{Qq}\propto f(x,Q,q)>0$.
The specific heat for CC ensemble is defined as
$C_{Qq}=T\left(\frac{\partial S}{\partial T}\right)_{Qq}
=-b\left(\frac{\partial S}{\partial b}\right)_{Qq}$.
So for the specific capacity heat to be positive,
$\left(\frac{\partial b}{\partial x}\right)_{Qq}<0$ should be satisfied which
is the result of the stability condition. Similarly, for the other ensembles,
we have 
\begin{eqnarray}
  \left(\frac{\partial I_{GC}}{\partial x}\right)_{Q\vphib}
  = (\bar b-b)\frac 1 b\left(\frac{\partial S}{\partial x}\right)_{Q\vphib}
  = -(\bar b-b)C_{Q\vphib}\left(\frac{\partial b}{\partial x}\right)_{Q\vphib}
  &=& 0 \qquad \textrm{(CG)} ;\nonumber\\
  \left(\frac{\partial I_{CG}}{\partial x}\right)_{\Phib q}
  = (\bar b-b)\frac 1 b\left(\frac{\partial S}{\partial x}\right)_{\Phib q}
  = -(\bar b-b)C_{\Phib q}\left(\frac{\partial b}{\partial x}\right)_{\Phib q}
  &=& 0 \qquad \textrm{(GC)} ;\nonumber\\
  \left(\frac{\partial I_{GG}}{\partial x}\right)_{\Phib\vphib}
  = (\bar b-b)\frac 1 b\left(\frac{\partial S}{\partial x}\right)_{\Phib\vphib}
  =(\bar b-b)C_{\Phib\vphib}\left(\frac{\partial b}{\partial x}\right)_{\Phi\vphib}
  &=& 0 \qquad \text{(GG)} ,
\end{eqnarray}
where the equilibrium conditions for $\varphi$ and $\Phi$ in
(\ref{eq:stationary-equation})  are used. Comparing with (\ref{eq:GC-first-condition}),
(\ref{eq:CG-first-condition}), (\ref{eq:GG-stationary-condition}),
and using (\ref{eq:CG-def-b-f}), (\ref{eq:GC-b-f}), and (\ref{eq:GG-func-b-f}), one
can find the equivalence of the positive specific heat capacity
conditions and our previous used conditions
(\ref{eq:GC-reduced-condition}), (\ref{eq:CG-reduced-condition}) and
(\ref{eq:GG-stability-condition}) for different ensembles,
respectively.

The positivity of the specific heat describes the stability of the system under the
fluctuations of horizon size or, equivalently, the temperature/entropy. In
\cite{Chamblin:1999hg,Braden:1990hw} the electric fluctuations are also considered
and the stability condition of one-charge black holes to these fluctuations depends
on the positivity of the electric permittivity. Similarly, here since the D$p$-D$(p+4)$
system has two kinds of charges, there can be two electrical stability conditions
for this system when these two kinds of electric fluctuations are considered.
According to the thermodynamics, the general stability conditions could be deduced
either from the maximal entropy or minimal energy criterion. As a result, one can
find out the requirements for different ``response functions'' suitable for the discussion
of stabilities in various ensembles,
\begin{eqnarray}
  C_{Qq} \equiv \left( \frac{\partial S}{\partial T} \right)_{Qq} > 0 ,\quad
  \epsilon_{TQ} = \left( \frac{\partial \varphi}{\partial q}\right)_{TQ} > 0 ,\quad
  {\cal E}_{T\varphi} = \left( \frac{\partial \Phi}{\partial Q}\right)_{T\Phi} > 0
  \qquad \text{(CC)} ;\nonumber\\
  C_{\Phi q} \equiv \left( \frac{\partial S}{\partial T} \right)_{\Phi q} > 0 ,\quad
  \epsilon_{T\Phi} = \left( \frac{\partial\varphi}{\partial q} \right)_{T\Phi} > 0 ,\quad
  {\cal E}_{Sq} = \left( \frac{\partial\Phi}{\partial Q}\right)_{Sq} > 0
  \qquad \text{(GC)} ;\nonumber\\
  C_{Q\varphi} \equiv \left( \frac{\partial S}{\partial T} \right)_{Q\varphi} > 0 ,\quad
  \epsilon_{SQ} = \left( \frac{\partial\varphi}{\partial q} \right)_{SQ} > 0 ,\quad
  {\cal E}_{T\varphi} = \left( \frac{\partial\Phi}{\partial Q} \right)_{T\varphi} > 0
  \qquad \text{(CG)} ;\nonumber\\
  C_{\Phi\varphi} \equiv \left( \frac{\partial S}{\partial T} \right)_{\Phib\varphi} > 0 ,\quad
  \epsilon_{S\Phi} = \left( \frac{\partial\varphi}{\partial q} \right)_{S\Phi} > 0 ,\quad
  {\cal E}_{Sq} =\left( \frac{\partial\Phi}{\partial Q} \right)_{Sq} > 0
  \qquad \text{(GG)} .
\end{eqnarray}
If there is only D$p$ brane charge, the positivity of $\cal E$ is  not
needed. In \cite{lu:2011-2}, in the grand canonical ensemble for
one-charge black brane, the electric stability condition is consistent with the
thermal stability condition of positive specific heat. However, this
is no longer
true in our case if one of the \Dp- or \Dppf-branes (or both of them) is in
canonical ensemble. The electrical stability analysis will  reshape the
phase structures in GC/CG/CC ensembles  and the branch of
the smaller black brane  will be unstable in all possible cases when the van der Waals-like phase
transition is supposed to happen. This is similar to the result in
\cite{Chamblin:1999hg}, where part of the smaller AdS black hole branch including the
van der Waals phase transition point is unstable when electrical
stability condition is considered.   After all, the brane system
is not so simple as the liquid-gas system because carrying charges is the nature
of branes and thus the electrical stability is almost inevitably as important
as its thermal stability. The complete analysis on the electric stability
of this system is a little complicated and we shall defer the details to a companion
paper \cite{xiao:2015}.

\section{Conclusions and outlook\label{sect:Conclude}}

In this paper, we have discussed different thermodynamical ensembles
of D$p$-D$(p+4)$ system, where $p=0,1,2$. The two kinds of charges can
be in either canonical or grand canonical ensemble separately, so there can
be CC, GG, GC, CG ensembles. CC ensemble has already been discussed in
\cite{lu:2012-2} and we focus on the other three ensembles in this
paper. In the following we summarize the results obtained in this
paper.

For GG ensemble, the potentials for D$p$ and D$(p+4)$ are fixed. At
very low temperatures, the hot flat space is the stable phase. Depending on
the values of $\bar \Phi$ and $\bar \varphi$, at higher temperatures
black brane phase may or may not be a globally stable phase.  For
larger $\bar \Phi$ or $\bar \varphi$ which satisfies
(\ref{eq:GG-Phi-phi-1}), black brane can not be a globally stable phase.
In this case, below $T_{\rm unstable}$ the globally stable phase is the hot flat space and
above it the horizon of  the system approaches the boundary and the
quantum effect will be important, therefore we do not know what
happens in this system under these circumstances. In this case, the system
is regarded as unstable in our semi-classical approach.
 Only for small
$\bar \Phi$ and $\bar \varphi$ satisfying (\ref{eq:GG-Phi-phi-2}), the
black brane can be a globally stable phase at temperature $T\in (T_{0},1/b(x_{max}))$.
At much higher temperatures,  like in case A, the horizon of the
system also tends to the boundary and the black brane is unstable.  As in most
grand canonical systems, there is no van der Waals-like phase transition.

For GC ensemble, in which the D$(p+4)$ charge $q$ and D$p$ potential
$\bar \Phi$ are fixed, the D2-D6, D1-D5 and  D0-D4 behave differently.
In D2-D6 system, on the right side of the dashed line 
in Figure~\ref{fig:GC-diagram-p-2},  the global minimum of
the reduced action is either at $x=q$ or at $x=x_{max}$, and the
black brane can not be a stable phase.  
On the left of the dashed line, in a comparably
small region of $\bar \Phi$, the black brane phase can be the final
stable phase only when the temperature is in range $(T_{\rm unstable},1/b_{GC}(x_{max}))$.
Below or above this range, the global minimum is at $x=q$ or $x=x_{max}$,
respectively, and the system is thus unstable.
For D1-D5 system, in the region below $q_c=1/3$ in Figure~\ref{fig:GC-diagram-p-2},
the discussion is similar to the D2-D6 system. For the region above $q_c$,
the black brane is always the final stable phase for temperature lower
than $1/b_{{GC}}(x_{max})$ and higher than $1/b_{{GC}}(q)$. For higher or lower
temperatures, the global minimum for the reduced action is at $x=x_{max}$
or $x=q$ which means the instability of the system. For D0-D4 system,
there can be a van der Waals-like first order phase transition between a
small black brane and a larger one in the region on the left of the dashed line in
region  B in Figure~\ref{fig:GC-diagram-p-0}, which is below the
critical charge $q_c$ at which a second order phase transition
happens. The critical charge is independent of $\bar
\Phi$ and is the same as the one for the black D4-brane in the canonical
ensemble. 
Below the first order phase transition temperature $T_t$  (or above $T_t$ and
below $1/b(x_{max})$), the final system is in a small black brane
phase (or a large one). For higher temperatures, the global minimum
is at $x=x_{max}$ again, which means that the system is unstable
and is beyond our approach to handle. In the region on the right of
the dashed line in B, or in region A, for temperatures below $T_{\rm unstable}$ or $1/b(x_{max})$,
respectively, the black brane is always the final stable phase, though there
may be a larger metastable one in the corresponding part in region B
for some temperature $T\in (1/b_{max},1/b(x_{max}))$. In region C, for
temperatures lower than $1/b(x_{max})$, the final stable phase is
always the black brane.

The fact that the critical charge $q_c$ in D0-D4 GC ensemble is independent
of $\bar \Phi$  is an unexpected result. Notice that according to
(\ref{eq:GC-Q-x}), $Q_{GC}$ still depends on $\bar \Phi$
non-trivially. Recall that for extremal D$p$-D$(p+4)$ branes
satisfying the Harmonic function rules and preserving 1/4
supersymmetries, there is no binding energy\cite{Tseytlin:1996hi},
which means that the two kinds of  charges do not affect each other.
However, we are considering non-extremal cases here, and the two kinds
of charges must be correlated, which is already demonstrated in the CC
ensemble where the critial line correlates both charges as in
Figure~\ref{fig:CC-critical-p01}.  Whether this independence of $\bar
\Phi$ of the critical charge in GC ensemble means that something
special happens in such a critial condition is an interesting problem
for future work.
 
From above discussion, we can see that D2-D6 GC ensemble behaves more
like GG ensemble. The $b_{GC}$ and $b_{GG}$ have the same behavior as
described by Figure~\ref{fig:GG-typical} and changes similarly as we
tune the $\bar \Phi$ in both GC ensemble and GG ensemble at small
$\bar \varphi$. The difference is that at $x=0$ in grand canonical
ensemble there is the hot flat space but at $x=q$ in GC ensemble the
extremal black brane has naked singularity and our method fails. There
could be new phases emerging around $x=q$ in GC ensemble like hot flat
space in GG ensemble due to quantum effects. D0-D4 GC ensemble is more
like the canonical ensemble. There can be a van der Waals-like phase
transition and there is a critical charge.  Tuning $q$ is similar to
that in the one-charge black brane canonical ensemble for small $\bar
\Phi$. But for large $\bar \Phi$ the van der Waals-like phase
transition may disappear.

In this paper, we only consider the thermal stability condition, i.e.
the positivity of the specific heat. The phase structure for CG ensemble
is almost the same as the one for GC ensemble except for the interchange in
$q\leftrightarrow Q$ and $\Phib\leftrightarrow \vphib$. Similar to CC
ensemble \cite{lu:2012-2}, the phase structure of GG ensemble is already
symmetric itself under this interchange. So the smeared D$p$ charges and
the D$(p+4)$ charges in the D$p$-D$(p+4)$ system are equipotent at least
so far as only the thermal stability is concerned. We will show explicitly
in another paper \cite{xiao:2015} that the electrical stability condition
will still preserve this symmetry, though more constraints shall be imposed.
However, we will show there that the phase structure will also be 
modified just like the situation which has been handled in \cite{Chamblin:1999hg}
for charged AdS black holes.

\appendix
\appendixpage

\section{Phase structures in CC ensemble}
\label{ap:CC}

\begin{figure}[t]
  \centering
  \includegraphics[width=.4\textwidth]{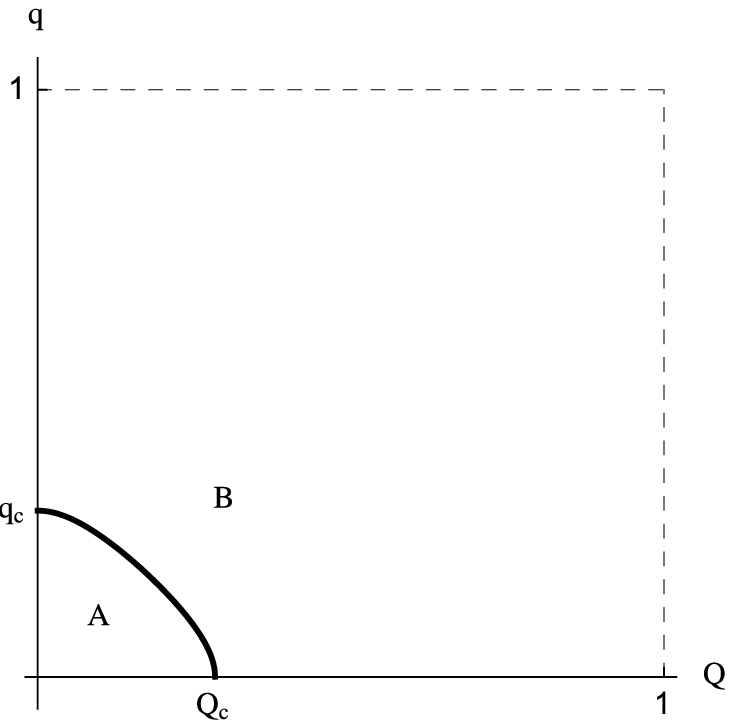} \qquad
  \includegraphics[width=.4\textwidth]{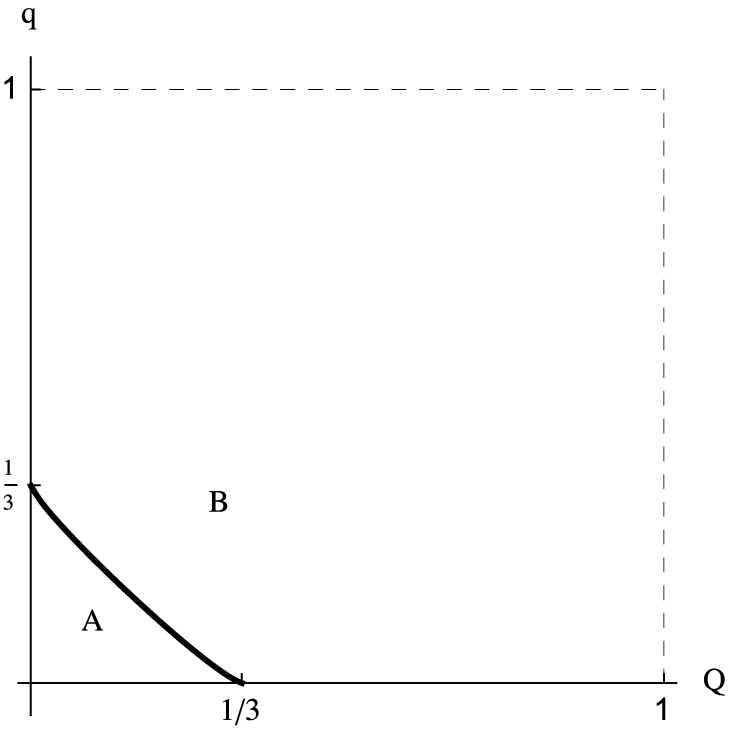}
  \caption{Critical lines in CC ensemble. The left diagram corresponds
    to D0-D4 case and the right one corresponds to D1-D5 case. The right 
    diagram has already been shown in \cite{lu:2012-2}. Both critical
  lines are computed numerically.}
  \label{fig:CC-critical-p01}
\end{figure}

Compared with the other ensembles, the analysis in CC ensemble is much simpler
due to the fact that both charges $Q$ and $q$ are fixed, which leads to
the consequence that
\begin{eqnarray}
  b_{CC}(1) = 0 .
  \label{eq:bCC-x-1}
\end{eqnarray}
Hence the shape of Figure~\ref{fig:GC-p-0} (a) would never occur, i.e.,
systems that have van der Waals-like structures can only have $b_{CC}(x)$
with shapes shown in Figure~\ref{fig:GC-p-0} (b) and (c). Therefore,
there is only one transition line in the $Q-q$ plane which is also
the critical line on which a second order phase transition occurs.
However, unlike in GC or CG ensemble where only D0-D4 can have van der
Waals-like phase transitions, in CC ensemble, both D0-D4 and D1-D5 can
have this behavior. In Figure~\ref{fig:CC-critical-p01} we show the
critical lines in $Q-q$ parameter plane. In both diagrams, B
corresponds to the region where $b_{CC}(x)$ is a monotonically
decreasing function and A corresponds to the region where van der
Waals-like phase transition can happen. In D0-D4 case, the critical
line intersects with the coordinate axes at $Q_c=q_c=0.141626$, which
exactly matches the result in the presence of only \Dp or \Dppf\
charges \cite{lu:2011}.

\begin{figure}[t]
  \centering
  \includegraphics[width=.45\textwidth]{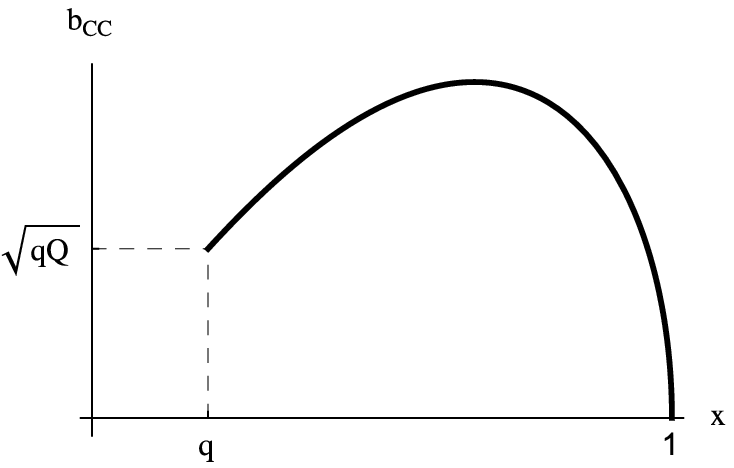} \quad
  \includegraphics[width=.45\textwidth]{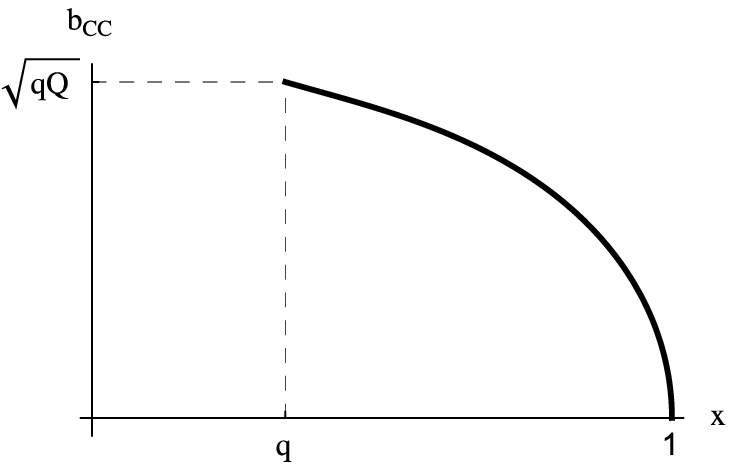}
  \caption{Possible shapes of $b_{CC}(x)$ for D2-D6-brane in CC ensemble.}
  \label{fig:CC-p2-shape}
\end{figure}

\begin{figure}[t]
  \centering
  \includegraphics[width=.5\textwidth]{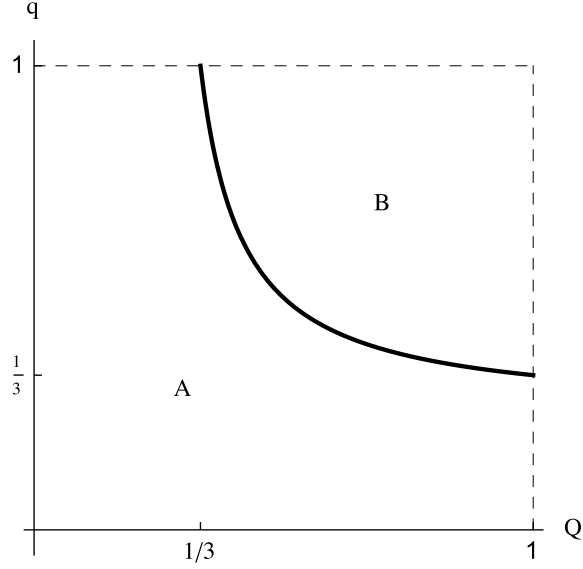}
  \caption{$Q-q$ plane for D2-D6-brane in CC ensemble.}
  \label{fig:CC-p2}
\end{figure}

For D2-D6 system, there are no critical behaviors. However, depending on
where $(Q,q)$ pair lies in the $Q-q$ plane there still can be two
possible shapes of $b_{CC}(x)$, as shown in Figure~\ref{fig:CC-p2-shape}.
The diagram for $Q-q$ plane is shown in Figure~\ref{fig:CC-p2}, in which
region A corresponds to the left diagram in Figure~\ref{fig:CC-p2-shape}
and region B corresponds to the right one. The boundary line between
these two regions can be obtained analytically to be
\begin{eqnarray}
  q = \frac{Q}{4Q-1} \qquad {\rm or} \qquad Q=\frac{q}{4q-1} ,
  \label{eq:CC-D2D6-line}
\end{eqnarray}
where $\frac13 < Q,q < 1$, which shows the symmetry between charge $Q$ and $q$.

\section{The condition for  globally stable black brane phases in the GG ensemble}
\label{ap:GG}

When $x=\bar{x}$ we have
\begin{eqnarray}
  \RIGG = \frac{5-p}{3-p} \left( \frac{\bar{x}}{1-\Phib^2} \right)^{1/2}
  \frac{(1-\bar\xi)^{\frac{1}{2}+\frac{1}{3-p}}}{1+\bar\xi^{1/2}}
  \left( \bar\xi^{1/2} - \frac{3-p}{5-p} \right) ,
  \label{eq:I-GG-xb}
\end{eqnarray}
where $\bar\xi=1-\bar{x}(1-\vphib^2)$. Now we need to find out when $\RIGG<0$,
or when
\begin{eqnarray}
  \bar\xi < \left( \frac{3-p}{5-p} \right)^2 .
  \label{eq:I-GG<0}
\end{eqnarray}
This leads to the condition that
\begin{eqnarray}
  \bar{x} > \bar{x}_0(\vphib) = \frac{4(4-p)}{(5-p)^2} \frac{1}{1-\vphib^2}
  \quad \textrm{or} \quad
  \bar{b} < b_{GG} \big( \bar{x}_0(\vphib) \big) .
  \label{eq:ap-GG-x0}
\end{eqnarray}
However, we have a restriction for $\bar{x}$ which is $x_0<\bar{x}<x_{max}$.
That means only if $\bar{x}_0(\vphib)<x_{max}$ we can have $\bar{x}>\bar{x}_0(\vphib)$.
It is easy to check that $\bar{x}_0(\vphib)>x_0$ is automatically satisfied, whereas
for $\bar{x}_0(\vphib)<x_{max}$ to hold we need
\begin{eqnarray}
  \textrm{max}\{\Phib,\vphib\} < \frac{3-p}{5-p} .
  \label{eq:ap-GG-xb0<xmax}
\end{eqnarray}
Thus we find the condition for the black brane phase to be globally stable,
\begin{eqnarray}
  \textrm{max}\{\Phib,\vphib\} < \frac{3-p}{5-p}
  \quad \textrm{and} \quad
  \bar{b} < b_{GG} \big( \bar{x}_0(\vphib) \big) .
  \label{eq:ap-GG-final-condition}
\end{eqnarray}

\section{Parameter planes in GC ensemble}
\label{ap:GC}

We have argued in section \ref{sec:GC-phase} that we can find the
transition lines between region A and region B in Figure
\ref{fig:GC-diagram-p-2} for $p =2$ by setting the $b_{GC}$ curve at the right end point
to be flat. This can also be used in arbitrary $p$. That is,
\begin{eqnarray}
  g(x_{max},q) = 0 ,
  \label{eq:transition-line}
\end{eqnarray}
where $x_{max}$ satisfies following equation (see (\ref{eq:GC-x-max})),
\begin{eqnarray}
  x_{max}^2 - (1-\Phib^2)x_{max} - q^2\Phib^2 = 0 .
  \label{eq:x-max-equation}
\end{eqnarray}
The above quadratic equation has two solutions and $x_{max}$ is the
larger one. Therefore, there is another restriction on $x_{max}$,
\begin{eqnarray}
  x_{max} > \frac{1-\Phib^2}{2}
  \label{eq:x-max-restriction}
\end{eqnarray}
From (\ref{eq:x-max-equation}), we can express $q$ through $x_{max}$,
\begin{eqnarray}
  q^2 = \frac{x_{max}^2 - (1-\Phib^2)x_{max}}{\Phib^2} .
  \label{eq:q-x-max}
\end{eqnarray}
Putting this relation into (\ref{eq:transition-line}), we can factorize
$g(x_{max},q)$ as follows,
\begin{eqnarray}
  g(x_{max},q) &=& \frac{x_{max}^2(1-x_{max})^2}{\displaystyle 2\Phib^4}
  \Big\{ \left[ 3(3-p)-(7-p)\bar\Phi^2 \right] x_{max} \nonumber\\
  && - 2(3-p-\Phib^2)(1-\Phib^2) \Big\} .
  \label{eq:transition-equation}
\end{eqnarray}
Thus we find the relation between $x_{max}$ and $\Phib$ when
(\ref{eq:transition-line}) is satisfied,
\begin{eqnarray}
  x_{max} = \frac{2(3-p-\Phib^2)(1-\Phib^2)}{\displaystyle 3(3-p)-(7-p)\Phib^2} .
  \label{eq:x-max-Phi}
\end{eqnarray}
Combining
(\ref{eq:q-x-max}) and (\ref{eq:x-max-Phi}), we finally reaches the
relation between $q$ and $\Phib$,
\begin{eqnarray}
  q = \frac{1-\Phib^2}{\Phib \big|3(3-p)-(7-p)\Phib^2\big|}
  \sqrt{2(3-p-\Phib^2)[(5-p)\Phib^2-(3-p)]} .
  \label{eq:transition-q-Phi}
\end{eqnarray}
From (\ref{eq:x-max-Phi})
and (\ref{eq:x-max-restriction}), the restriction
on $x_{max}$ becomes a restriction on $\Phib$,
\begin{eqnarray}
  \Phib < \sqrt{\frac{3(3-p)}{7-p}} .
  \label{eq:Phi-restriction}
\end{eqnarray}
For $p=0$ or 1, this restriction is satisfied automatically since $\Phib<1$
while for $p=2$, this inequality reduces to $\Phib<\sqrt{3/5}$. 
Now it
is easy to see that, for $p=0$ or 1, (\ref{eq:transition-q-Phi}) reduces to
(\ref{eq:GC-boundary-p-0}) or (\ref{eq:GC-boundary-p-1}) and for
$p=2$ together with (\ref{eq:Phi-restriction}) this recovers
(\ref{eq:GC-boundary-p-2}).
 
For $p=1$, there is another transition line between region B and
region C in Figure~\ref{fig:GC-diagram-p-1} which appears when the
$b_{GC}$ curve becomes flat at the left end point. That is equivalent
to setting
\begin{eqnarray}
  \DD{b_{GC}(q)}{x} = - \frac{(q-1/3)^2}{q^2(1-q)^2} b_{GC}(q) = 0 .
  \label{eq:transition-p-1}
\end{eqnarray}
This gives a condition which is independent of $\Phib$,
\begin{eqnarray}
  q = 1/3 .
  \label{eq:transiton-q-p-1}
\end{eqnarray}
So the transition line is the horizontal line shown in Figure
\ref{fig:GC-diagram-p-1}.

For $p=0$, there is a critical line between region A and C  on which
the $b_{GC}$ has a inflection point and the first and second
order derivatives of $b_{GC}$ vanishes at the same point $\bar{x}$.
That means, we only need to combine two equations,
\begin{eqnarray}
  g(\xb,q) = 0 ,\nonumber\\
  \PP{g(\xb,q)}{x} = 0 .
  \label{eq:critical-line}
\end{eqnarray}
This set of equations are solved by
\begin{eqnarray}
  \xb \cong 0.292675 ,\quad q_c \cong 0.141626 .
  \label{eq:critical-x-q}
\end{eqnarray}
Since $x<x_{max}$, this critical point exists only when $x_{max}>\xb$,
which means there is an upper bound for $\Phib$,
\begin{eqnarray}
  \Phib_{max} \cong 0.871417 .
  \label{eq:Phi-upper-bound}
\end{eqnarray}

\section{Parameter planes in CG ensemble}
\label{ap:CG}

We have argued that the behavior of $b_{CG}(x)$ depends on the signature
of $b_{CG}^{\,'}(x)$. Hence we need to find the zero points of
$b_{CG}^{\,'}(x)$. With (\ref{eq:func-q-x}), one can easily find
following relations,
\begin{eqnarray}
  \BDM &=& \frac{\BDP}{\xi} \nonumber,\\
  \BDS &=& \frac{\lambda}{\xi} \BDP ,
  \label{eq:ap-dm-ds}
\end{eqnarray}
where
\begin{eqnarray}
  \xi &=& 1 - (1-\bar{\varphi}^2)x ,\nonumber\\
  \lambda &=& \frac{1+\xi+\sqrt{(1-\xi)^2+4Q^2\xi}}{2(1-Q^2)} .
  \label{eq:ap-def-xi-lambda}
\end{eqnarray}
With this relation one can evaluate $b_{CG}^{\,'}(x)$ and set it to
zero,
\begin{eqnarray}
  b_{CG}^{\,'}(x) = \frac{b_{CG}}{2(3-p)x}
  \frac{\xi\big[3(3-p)-(7-p)\xi\big] - \lambda(\xi+1)(3-p-2\xi)}{\xi(\xi\lambda+\lambda-2\xi)} = 0 ,
  \label{eq:ap-db-dx}
\end{eqnarray}
which results in,
\begin{eqnarray}
  \lambda &=& \frac{\xi\big[3(3-p)-(7-p)\xi\big]}{(\xi+1)(3-p-2\xi)} .
  \label{eq:ap-lambda}
\end{eqnarray}
The second definition in (\ref{eq:ap-def-xi-lambda}) is equivalent to
requiring
\begin{eqnarray}
  0 &=& (1-Q^2)\lambda^2 - (1+\xi)\lambda + \xi ,\quad \text { and
}\quad
  \lambda > \frac{1+\xi}{2(1-Q^2)} ,
  \label{eq:lambda-equation}
\end{eqnarray}
by the same token of (\ref{eq:x-max-equation}) and
(\ref{eq:x-max-restriction}). Combining (\ref{eq:ap-lambda}) and
(\ref{eq:lambda-equation}), one gets
\begin{eqnarray}
  Q &=& \frac{1-\xi}{3(3-p)-(7-p)\xi}
  \sqrt{\frac{2(3-p-\xi)[(5-p)\xi-(3-p)]}{\xi}} ,\nonumber\\
  \xi &<& \left\{ \begin{array}{ll}
    1 , & {\rm for\ } p=0,1 \\
    \frac{3}{5} , & {\rm for\ } p=2
  \end{array} \right. .
  \label{eq:relation-Q-xi}
\end{eqnarray}

By the same spirit of the previous appendix section, we examine when the slope
at each end point of $b_{CG}$ curve changes its signature. That is to
set $x=0$ and $x=1$ for left and right end points respectively. For 
$x=0$ to be the stationary point, or equivalently at $\xi=1$, which
only makes sense when $p=1$ (because only this case we have finite
limit of $b_{CG}$), we
have $Q=1/3$. This corresponds to the horizontal line in
Figure~\ref{fig:CG-diagram-p-1}. For $x=1$, or equivalently $\xi=\vphib^2$,
we get
\begin{eqnarray}
  Q &=& \frac{1-\vphib^2}{3(3-p)-(7-p)\vphib^2}
  \sqrt{\frac{2(3-p-\vphib^2)\big[(5-p)\vphib^2-(3-p)\big]}{\vphib^2}}
  \label{eq:transition-Q-phi}
\end{eqnarray}
and $\vphib<\sqrt{\frac{3}{5}}$ for $p=2$. This relation represents all
curves in Figure~\ref{fig:CG-diagram-p-2}, \ref{fig:CG-diagram-p-1}
and \ref{fig:CG-diagram-p-0}.

Finally, we calculate the critical charge $Q_c$ for the $p=0$ case.
From (\ref{eq:relation-Q-xi}) we know that
\begin{eqnarray}
  Q_c = \frac{1-\xi}{9-7\xi} \sqrt{\frac{2(3-\xi)(5\xi-3)}{\xi}} .
  \label{eq:Q-c-equation-1}
\end{eqnarray}
Another critical condition is that $b_{CG}^{\,''}(x)=0$. Since
\begin{eqnarray}
  b_{CG}^{\,''}(x) \eval{b_{CG}^{\,'}(x)=0}
  &\sim& (4\xi-1)\lambda - (14\xi-9) +
  \frac{(\xi+1)(2\xi-3)(\lambda-1)}{2(1-Q_c^2)\lambda-1-\xi} ,
  \label{eq:ap-d2b-dx2}
\end{eqnarray}
we find another equation for $Q_c$,
\begin{eqnarray}
  2(1-Q_c^2) &=& \frac{(\xi+1)^2(2\xi-3)}{\xi(7\xi-9)} \left[ 1 +
  \frac{(\xi-1)(2\xi-3)(5\xi-3)}{(\xi-3)(11\xi-9)} \right] .
  \label{eq:Q-c-equation-2}
\end{eqnarray}
Combining (\ref{eq:Q-c-equation-1}) and (\ref{eq:Q-c-equation-2}),
we get only one sensible solution
\begin{eqnarray}
  \xi_c \cong 0.759367 ,\quad Q_c \cong 0.141626 .
  \label{eq:Q-c-solution}
\end{eqnarray}
This also says that, on the critical line, $x$ and $\bar{\varphi}$
are related through a simple formula,
\begin{eqnarray}
  x = \frac{1-\xi_c}{1-\vphib^2} \cong \frac{0.240633}{1-\vphib^2} .
  \label{eq:ap-rel-x-phi}
\end{eqnarray}
Since $x<1$, this formula also indicates that $\vphib$ has a maximal
value beyond which $Q_c$ does not exist, i.e.,
$\vphib_{max}=\sqrt{\xi_c}\cong 0.871417$.

\section{$b_{\rm unstable}$ in GC ensemble and the dashed lines in Figure~\ref{fig:GC-diagram-p-2} and \ref{fig:GC-diagram-p-1} }
\label{ap:dashed-lines}

Our starting point is to find the critical $\bar{b}=b_{\rm unstable}$
above which we have $I_{GC}(q)<I_{GC}(\bar{x})$ and the system at
$x=\bar{x}$ is not globally stable. As a bonus of this process we
will find the dashed lines in Figure~\ref{fig:GC-diagram-p-2} and
\ref{fig:GC-diagram-p-1}. When $\bar{b}=b_{\rm unstable}$ we have
\begin{eqnarray}
  \RIGC(q) = \RIGC(\bar{x})
  \label{eq:I-q-eq-I-xb}
\end{eqnarray}
where $\bar{x}$ takes the value such that
\begin{eqnarray}
  \bar{b} = b(\bar{x}) .
  \label{eq:bb-eq-b-xb}
\end{eqnarray}
In (\ref{eq:I-q-eq-I-xb}) the left and right hand side can be evaluated,
\begin{eqnarray}
  \RIGC(q) &=& (3-p) \bar{b} q ,\nonumber\\
  \RIGC(\bar{x}) &=&  \bar{b} \left[ 8-2p - (5-p) \left( \frac{\BDP(\bar{x})}{\BDM(\bar{x})}
    \right)^{1/2} - (3-p) \left( \BDP(\bar{x}) \BDM(\bar{x}) \right)^{1/2} \right] \nonumber\\
  && - \left( \frac{\bar{x}}{1-\Phib^2} \right)^{1/2}
    \left( 1-\frac{\BDP(\bar{x})}{\BDM(\bar{x})} \right)^{3/2} \nonumber\\
  &=& \bar{b} \Bigg[ 8-2p - 2 \left[ \frac{\BDP(\bar{x})}{\BDM(\bar{x})}
    \right]^{\frac{1}{2}} - (3-p) \left[ \BDP(\bar{x}) \BDM(\bar{x}) \right]^{\frac{1}{2}}
    - (3-p) \left[ \frac{\BDM(\bar{x})}{\BDP(\bar{x})} \right]^{\frac{1}{2}} \Bigg] \qquad
  \label{eq:action-evaluated}
\end{eqnarray}
where we have used (\ref{eq:other-reduce-action}) (\ref{eq:GC-delta-star-x}) (\ref{eq:GC-Q-x})
and (\ref{eq:bb-eq-b-xb}) to obtain the final expressions. Combining
(\ref{eq:I-q-eq-I-xb}) and (\ref{eq:action-evaluated}) we get an equation
of $\bar{x}$,
\begin{eqnarray}
  8 - 2p - 2 \left[ \frac{\BDP(\bar{x})}{\BDM(\bar{x})}
  \right]^{\frac{1}{2}} - (3-p) \left[ \BDP(\bar{x}) \BDM(\bar{x}) \right]^{\frac{1}{2}}
  - (3-p) \left[ \frac{\BDM(\bar{x})}{\BDP(\bar{x})} \right]^{\frac{1}{2}}
  = (3-p) q .
  \label{eq:equation-xb}
\end{eqnarray}
That is, given any pair of $(\Phib,q)$ we can solve (at least numerically)
the above equation to find the $\bar{x}$ that makes $\RIGC(q)$ and
$\RIGC(\bar{x})$ equal. One can easily see that $\bar{x}=q$ is always
a solution to this equation which is not the solution we want. However,
it can be shown that if there exists a solution other than $q$, this
solution is unique and therefore there exists a unique $b_{\rm unstable}=b(\bar{x})$.
Consequently, there may be regions and boundaries thereof in the
$q$-$\Phib$ plane within which this solution always exists. These regions
should lie in region B of Figure~\ref{fig:GC-diagram-p-2} and \ref{fig:GC-diagram-p-1}
and the boundaries are exactly the dashed lines in these figures.
We have argued that there is an upper limit for $x$ which is $x_{max}$ defined
in (\ref{eq:GC-x-max}), therefore on the boundary we have $\bar{x}=x_{max}$.
Bringing this specific value of $\bar{x}$ in (\ref{eq:equation-xb}), we find
\begin{eqnarray}
  q = \frac{2}{3-p} \frac{(1-\Phib)(3-p-\Phib)[(5-p)\Phib-(3-p)]}{\Phib[3(3-p)-(7-p)\Phib]} .
  \label{eq:q-phi-dashed-line}
\end{eqnarray}
Setting $p=2$ and $p=1$ respectively the expected functions in
(\ref{eq:q-Phi-p-2}) and (\ref{eq:q-Phi-p-1}) could be recovered.

\acknowledgments

This work is supported by the National Natural Science Foundation of
China under grant No. 11105138, and 11235010. Z.X is also partly
supported by the Fundamental Research Funds for the Central
Universities under grant No.~WK2030040020. D.Z is partly supported
by the Chinese Scholarship Council. He would also like to thank
Chao Wu, Wei Gu and Jianfei Xu for their helpful discussions as
well as Prof.~Yang-Hui He and City University London for providing
him with a one-year visiting studentship.

\end{document}